\documentclass[aps,reprint,amsmath,amssymb,prb]{revtex4-1}

\usepackage{xcolor}
\usepackage{graphicx} 
\usepackage{dcolumn} 
\usepackage{amsmath} 
\usepackage[version=4]{mhchem}

\def\approxprop{%
  \def\p{%
    \setbox0=\vbox{\hbox{$\propto$}}%
    \ht0=0.6ex \box0 }%
  \def\s{%
    \vbox{\hbox{$\sim$}}%
  }%
  \mathrel{\raisebox{0.7ex}{%
      \mbox{$\underset{\s}{\p}$}%
    }}%
}

\begin{document}

\title{Temperature Study of Rydberg Exciton Optical Properties in \ce{Cu2O}
}

\author{Dongyeon Daniel Kang$^{1,2\dagger}$, Aaron Gross$^{1,2\dagger}$, HeeBong Yang$^{1,2\dagger}$, Yusuke Morita$^{3}$,  Kyung Soo Choi$^{1,5,7}$, Kosuke Yoshioka$^{4}$, Na Young Kim$^{1,2,5,6,7*}$}
\address{$^1$Institute for Quantum Computing,  University of Waterloo, 200 University Ave. West, Waterloo, ON, N2L 3G1, Canada}
\address{$^2$Department of Electrical and Computer Engineering,  University of Waterloo, 200 University Ave. West, Waterloo, ON, N2L 3G1, Canada}
\address{$^3$Department of Physics, Graduate School of Science, The University of Tokyo, 7-3-1 Hongo, Bunkyo-ku, Tokyo, 113-0033, Japan}
\address{$^4$Photon Science Center, School of Engineering, The University of Tokyo, 2-11-16 Yayoi, Bunkyo-ku, Tokyo, 113-8656, Japan}
\address{$^5$Department of Physics and Astronomy,  University of Waterloo, 200 University Ave. West, Waterloo, ON, N2L 3G1, Canada}
\address{$^6$Department of Chemistry,  University of Waterloo, 200 University Ave. West, Waterloo, ON, N2L 3G1, Canada}
\address{$^7$Perimeter Institute for Theoretical Physics, 31 Caroline St. N, Waterloo, ON, N2L 2Y5, Canada}
\homepage{$\dagger$The authors equally contribute to the work.}
\email{$^*$Corresponding autor: nayoung.kim@uwaterloo.ca}

\date{\today}

\begin{abstract}
Rydberg excitons in \ce{Cu2O} can be an emergent platform for solid-state quantum information processing by utilizing the exaggerated properties of high-lying excited states within the material. To develop practical quantum systems, high-temperature operation is desirable. Here, we study the temperature-dependence of the yellow and green Rydberg exciton resonances in a thin \ce{Cu2O} crystal via broad-band phonon-assisted absorption spectra between 4 K and 100 K. 
At 4 K, we can identify the principal quantum number $n=11$ yellow and $n=4$ green Rydberg exciton states, beyond which we are limited by the spectral resolution of standard absorption techniques.
Above liquid nitrogen boiling temperature ($\sim$80 K), the $n=6$ yellow and $n=4$ green Rydberg exciton states are readily captured 
and higher-temperature yellow Rydberg exciton optical properties still exhibit the standard scaling laws seen at low temperatures.
This promising result lays the groundwork for a new route to build a high-temperature Rydberg quantum information processing architecture with solid-state \ce{Cu2O}.
\end{abstract}
\maketitle 

\section{Introduction}

Rydberg atoms have been actively investigated for the purpose of building promising quantum information science and technologies based on their exaggerated optical properties stemming from high-lying excited states. These states typically have a large principal quantum number of $n >$ 100, leading to stronger interaction strength and higher sensitivity to environmental effects like electric fields \cite{Saffman2010, Adams2019}. One manifestation of these optical properties is the set of beautiful scaling laws of the binding energies, spectral linewidth, and oscillator strength with respect to principal quantum number $n$.
Similar to Rydberg atoms, Rydberg excitons refer to highly excited hydrogen-like quantum quasi-particles in solid-state systems. Their presence were seen in two solid-state systems: one is transitional dichalcogenide monolayers up to $n$ = 4 or 5 \cite{Chernikov2014,Stier2018}, and the other is cuprous oxide, \ce{Cu2O}. The previous record high of $n$ = 25 from the P-series yellow excitons in {\ce{Cu2O}}, observed via high-resolution absorption spectroscopy at 1.2 K {\cite{Kazimierczuk2014}}, was recently surpassed by an observation of $n$ = 28 at 760 mK {\cite{Heckotter2020_2}}.
Since 1952, when excitons in \ce{Cu2O} were first observed \cite{Hayashi1952a,Hayashi1952b}, the material has been at heart of both experimental and theoretical studies to explore fundamental physics and engineering applications \cite{Itoh1975,Elliott1957,Kubouchi2005, Yoshioka_2012,Snoke2014, Forman1971,Olbrychski1975,Eng_Ito1998,solar_Malerba2011}. Dipole-forbidden 1S-para excitons in \ce{Cu2O} are long-lived ($\sim$ 300-500 ns for a high-purity sample) and have become a promising solid-state candidate to realize excitonic Bose-Einstein condensation (BEC) \cite{Snoke2014}. Bulk \ce{Cu2O} has also been used to make low-cost photovoltaic cells to harness solar energy \cite{Abdu2009,solar_Malerba2011}.

The recent observation of giant Rydberg excitons and their optical properties from the yellow P-excitons in a thin \ce{Cu2O} crystal \cite{Kazimierczuk2014} has ignited new research activities including the exploration of a potential solid-state Rydberg system for quantum information processing \cite{Khazali2017,Heckoetter2020,Ziemkiewicz2020a,Ziemkiewicz2020b}. 
Although the highest $n$ of the Rydberg exciton is only 28, comparatively lower than the typical $n >$ 100 number of Rydberg atoms, at this lower principal quantum number Rydberg excitons exhibit compatible optical properties in terms of beautiful scaling laws similar to their atomic counterparts \cite{Kazimierczuk2014,Heckotter2017,Zielinska-Raczynska2016,Schweiner2017,Heckotter2017,Kitamura2017,Schone2017,Semina2018,Stolz2018,Takahata2018}: a micron-sized average exciton radius, a huge Blockade volume ($\sim$ 1000 $\mu$m$^3$) arising from a strong dipole-dipole interaction, and an increased lifetime of $\sim$ 1 ns, $\sim$ 1,500 times longer than the short lifetime of the 2P exciton state. 
J. Heck{\"{o}}tter \emph{et al.} thoroughly discussed the scaling law of the yellow excitons in \ce{Cu2O} and their responses to electric and magnetic fields at temperatures lower than 2 K \cite{Heckotter2017}; Takahata \emph{et al.} provided a complete photoluminescence (PL) spectroscopy and spectroscopic analysis, including temperature dependence of the energy shift of P-series of the yellow excitons between 6 and 100 K \cite{Takahata2018}. 

Indeed, Rydberg energies in \ce{Cu2O} of 80 - 150 meV, one-fold larger than those of typical semiconductors (e.g. GaAs $\sim$ 4 meV), can stabilize excitons against a thermal environment at higher temperatures. Here, we measure the optical properties of both yellow and green Rydberg excitons over a range of temperatures (4 - 100 K) followed by careful analysis in order to examine practical uses of \ce{Cu2O} Rydberg excitons at elevated temperatures in quantum applications. High-temperature operation is an attractive way to circumvent some of the technological barriers of other platforms, such as the necessity of dilution refrigeration in many solid-state quantum materials. Our analysis aims to cover a gap left in the research of this system, since Rydberg excitons in \ce{Cu2O} have so far been predominantly studied at low temperatures of around 1.5 K or below. 

According to our systematic and thorough analysis of Rydberg exciton resonances obtained at various temperatures above 4 K, we conclude that Rydberg excitons with $n \sim$ 8 - 10 are stable enough to show exaggerated Rydberg properties which could be utilized for high-temperature quantum information processing. Ultimately, we seek for a scalable room-temperature (RT) solid-state system to build quantum hardware. As a promising RT material, quantum defects in diamond and SiC have been demonstrated as foundations for forming a qubit and favorable RT experimental results are reported~\cite{Wrachtrup2010,Weber2010,Bassett2019}. However, successful coherent qubit operations of quantum defect qubits remain at low temperatures below 10 K, where the environmental thermal effect is reduced. In addition, the mutual interaction between quantum defects is natively weak. So far, there are currently no solid-state platforms capable of high-temperature operation for introducing long-range interaction. Hence, we envision that \ce{Cu2O} Rydberg excitons can offer a novel quantum Rydberg architecture at above liquid-nitrogen temperature.   

We organize our paper as follows: First, Sec. II summarizes the band structures of \ce{Cu2O} to introduce the yellow and green exciton series followed by our experimental setup and measurement methods. We present our phonon-assisted absorption spectra and results from our extensive analysis in Sec. III. We examine the scaling law behavior (Sec. IV) and their temperature dependence (Sec. V) of both yellow and green excitons. Sec. VI discusses the comparison of Rydberg excitons with Rydberg atoms in terms of their optical properties and potential future quantum information processing applications. 

\section{Band Structure and Experiments}

\subsection{Band Structure}

\ce{Cu2O} crystal has a cubic structure with full octahedral symmetry, whose basic bandstructures are formed by Cu 3$d$ and 4$s$ electron orbitals. As can be seen in Fig. \ref{Fig1:PhononAssistedAbsorption}(a), there are two lowest conduction bands with symmetry of $\mathrm{\Gamma_{6}^{+}}\otimes \mathrm{\Gamma_{8}^{-}}$ and two highest valence bands with $\mathrm{\Gamma_{7}^{+}}\otimes \mathrm{\Gamma_{8}^{+}}$ \cite{Schone2017}. \ce{Cu2O} is a direct bandgap semiconductor \cite{Kavoulakis1997}, where the yellow or green exciton transitions occur in the visible wavelength range (530 - 580 nm) between the $\mathrm{\Gamma_{6}^{+}}$
conduction band and  $\mathrm{\Gamma_{7}^{+}}$ or $\mathrm{\Gamma_{8}^{+}}$ valence bands which are separated by a spin-orbit interaction energy of 131 meV. There also exist well-known higher-energy series, the so-called blue or violet excitons, between the $\mathrm{\Gamma_{8}^{-}}$ conduction band and  $\mathrm{\Gamma_{7}^{+}}$ or $\mathrm{\Gamma_{8}^{+}}$ valence bands, respectively. Due to the same parity of the envelope functions, the lowest 1S excitons in the yellow and green series are dipole-forbidden but quadrupole allowed, while the P excitons are dipole allowed. Consequently, the P-excitons of both series are optically accessible, appearing in standard absorption spectra, which led to the observation of the yellow P-excitons with high numbers of $n$ up to 28 {\cite{Heckotter2020_2}}. This work focuses on both the yellow and green P-excitons.  

\subsection{Experimental Methods}

Natural crystalline \ce{Cu2O} from Namibia was cut and polished to be roughly 65 $\mu$m-thick. We placed this thin \ce{Cu2O} crystal in a strain-free sample holder inside a temperature controllable \textsuperscript{4}\ce{He} cryostat, and we measured transmission spectra covering the yellow and green exciton series at varying temperatures (4 K $< T <$ 100 K). An incident light was shined and focused onto the surface of the sample, with a broad-band white light (Ando AQ-4303B, power density of 150 $\mu$W/mm$^2$) for the yellow excitons and a green light emitting diode for the green excitons. 

We also took PL spectra with a green laser at 520 nm.  
In order to quantify physical parameters of exciton energies, binding energies, spectral linewidth, and oscillator strength, we collected and analyzed transmission signals with a monochromator (Princeton Instruments SP2750) whose focal length is 750 mm and spectral resolution is $\leq$ 80 $\mu$eV, equivalent to $\sim$20 pm or 19.3 GHz. 
Following a standard procedure, we normalize the measured spectra to have zero-absorption at an energy of 2.03 eV ($\sim$610 nm) that corresponds to the lowest energy of phonon-assisted absorption, effectively removing artificial loss like surface reflection and scattering. We obtain experimental absorption coefficient spectra ($\alpha (E, T)$) as a function of energy ($E$) at different temperatures ($T$) from the aforementioned transmission spectroscopy 
$$ \text{Transmission} \propto \exp(-\alpha (E,T) d_{\ce{Cu2O}})$$ 
by assuming the crystal thickness to be $d_{\ce{Cu2O}}$ = 65 $\mu$m. 

\section{Absorption spectroscopy for Rydberg exciton series in C\MakeLowercase{u}\textsubscript{2}O}

\begin{figure} 
\centering
\includegraphics[width=1.0\columnwidth]{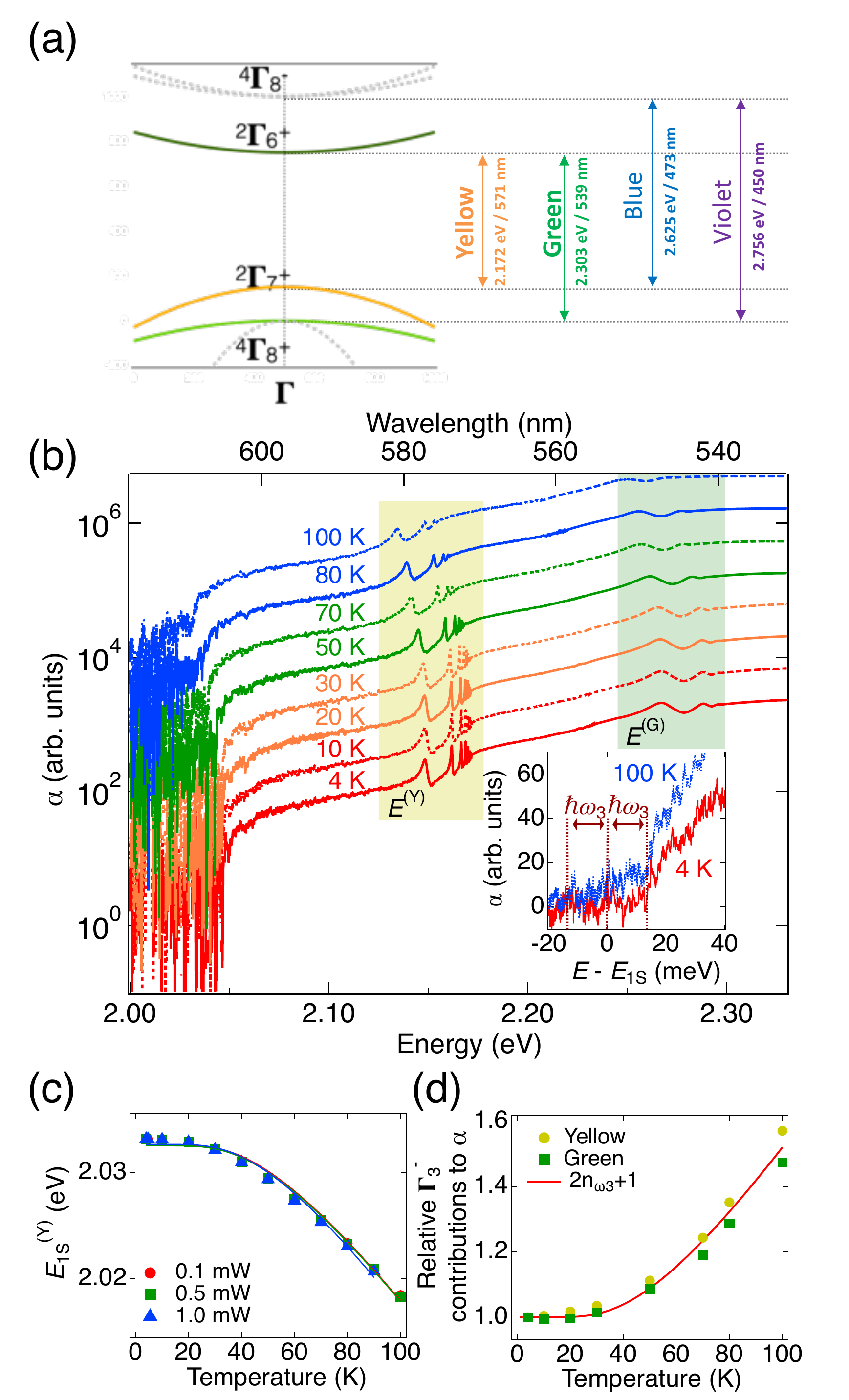}
\caption{Temperature-dependent phonon-assisted absorption spectra of \ce{Cu2O}. (\textbf{a}) A simple electronic bandstructure of \ce{Cu2O} exciton transitions at zone center. (\textbf{b}) Absorption coefficient spectra $\alpha$ taken from 4 K to 100 K, with a vertical shift to guide the eyes. The inset shows a close-up near the 1S ortho-exciton energy, and the dotted vertical lines indicate directly the $\Gamma_3^-$ energy, $\hbar\omega_3$ = $\pm$ 13.6 meV with respect to the energy shift. (\textbf{c}) The peak energy shift of the 1S-ortho yellow exciton as a function of temperature from separate photoluminescence measurements (Appendix C). Solid line are the best fits to Eq.(1) and the fitting coefficient values are enlisted in Table \ref{tab1:E1SCotFitRes}. (\textbf{d}) The estimated phonon-assisted absorption contribution from major ($\Gamma_{3}^{-}$) phonons associated with the 1S yellow and green excitons as a function of temperature. The solid line is the total phonon densities from phonon creation ($n_{pn}$) and destruction ($n_{pn}$+1) at a given temperature, where $n_{pn}$ is the Bose-Einstein distribution for $\Gamma_{3}^{-}$ phonon energy, $\hbar \omega_3$. 
}
\label{Fig1:PhononAssistedAbsorption}
\end{figure}

\subsection{Phonon-Assisted Absorption Spectra}

Figure \ref{Fig1:PhononAssistedAbsorption}(b) plots $\alpha (E, T) $ at eight different temperatures, where each spectrum is vertically shifted for clarity. The overall $\alpha (E, T) $ primarily consists of the environmental broad S-exciton absorption  $\alpha_{n\text{S}}$, the exponential Urbach tail and dipole-allowed sharp $n$P-exciton resonances both in yellow ($\alpha_{n\text{P}}^{\text{(Y)}}$, 570 - 580 nm) and green  ($\alpha_{n\text{P}}^{\text{(G)}}$, 530 - 550 nm) color regions.

\subsubsection{Thermal-Phonon Dependence}

First, the continuum part of $\alpha_{n\text{S}} (E, T)$ is associated with the phonon scattering processes of 1S-ortho exciton, 
which follows the well-known Elliott's formula, proportional to a quasi-square-root of the energy difference from the absorption origin energy, for example, 1S-exciton energy ($E_{1S}$) within the second-order perturbation theory \cite{Elliott1957}. 
In \ce{Cu2O}, ${\Gamma}_3^{-}$ phonon energy ($\hbar\omega_3$ = 13.6 meV) is most significant among other symmetricity-allowed phonons such as ${\Gamma}_4^{-}$ ($\hbar \omega_4$ = 82.1 meV) and minor phonons. The dominance of ${\Gamma}_3^{-}$ phonons is explicitly justified by two independent absorption and PL experimental results. The band edge of the absorption spectra at a particular temperature occurs exactly shifted by $\hbar\omega_3 =$ 13.6 meV (Fig. \ref{Fig1:PhononAssistedAbsorption}(b), inset). The ${\Gamma}_3^{-}$ phonon-associated 1S-ortho exciton PL peak also resides at the energy away from the 1S-ortho exciton PL peak by $\hbar\omega_3$ (Fig. \ref{Fig9:PL} in Appendix C).

\begin{table}
 \caption{Coefficients of the 1S-ortho exciton energy $E_{\text{1S}} (T)$ from photoluminescence spectra (Fig.\ref{Fig1:PhononAssistedAbsorption}(c)) between 4 K and 100 K, fitted with Eq. (\ref{eq:Tdep}).}
    \label{tab1:E1SCotFitRes}
    \begin{ruledtabular}
    \begin{tabular}{cccc} 
    Power (mW) & $E_{\text{1S,0}}$ (meV) & $a$ (meV) & $\Theta$ (K) \\  
 \hline 
        0.1 & 2048.13  & 15.06 & 112.33 \\ 
        0.5 & 2047.93  & 14.86 & 111.01\\ 
        1.0 & 2046.97  & 13.88 & 106.04 \\ 
        \hline
        average & 2047.68 $\pm$ 0.62 & 14.60 $\pm$ 0.63 & 109.79 $\pm$ 3.32 \\ 
   \end{tabular}
   \end{ruledtabular}
\end{table}

As temperature ($T$) increases, the band edge, the P-exciton resonances in the absorption spectra and the phonon-assisted 1S-exciton peaks in the PL are all progressively red-shifted. The 1S-exciton energy $E_{1S}(T)$, is plotted along $T$ in Fig. \ref{Fig1:PhononAssistedAbsorption}(c). This temperature dependence fits very well to the following equation (solid lines):

\begin{equation}\label{eq:Tdep}
E_\text{1S}(T) = E_\text{1S,0}- a \left(1+\frac{2}{e^{\Theta/T}-1}\right),
\end{equation}
where $E_\text{1S,0}$, $a$ and $\Theta$ are fitting coefficients. The fitting results from the PL data at three laser power values are in Table \ref{tab1:E1SCotFitRes}, which tells us that these power levels do not heat our sample above the cryostat temperature. 

As expected, $\Gamma_3^-$ phonon contributions get stronger for higher temperatures, which is explicitly visible in Fig. \ref{Fig1:PhononAssistedAbsorption}(d). We evaluate them from the ratio of the absorption coefficient values $\alpha_{n\text{S}}$ at each temperature with respect to that of 4 K by integrating the spectra in the energy window ($E-E_\text{1S} \in \{0.01, 0.07\}$ eV) for the yellow and  ($E-E_\text{1S} \in \{0.14, 0.18\}$ eV) for the green excitons. The data perfectly follows 2$n_{\omega_3}$+1, the sum of both phonon creation and destruction processes neglecting the phonon-energy shift without any fitting parameters, where $n_{\omega_3}$ is the $\Gamma_3^-$ phonon density in the form of a Bose-Einstein distribution,
\begin{equation*}
n_{\omega_3}(T)=\frac{1}{\exp \left(\hbar \omega_3 / k_B T\right)-1}, \end{equation*}
where $\hbar$ is the Planck constant $h$ divided by 2$\pi$ and $k_B$ is a Boltzmann constant.
In addition to the continuum thermal phonon background, the P-exciton absorption spectra still contains an Urbach tail due to material quality near the band edge and distinctive sharp peaks at optical dipole resonance transitions. Figure \ref{Fig7:absco} in Appendix A depicts major and minor phonon-assisted contributions to the broad background signal.

\subsubsection{Rydberg Exciton Resonances}

After subtracting the continuum thermal phonon background and Urbach tail, the remaining part of the P-exciton absorption spectra is series of distinctive sharp peaks attributed to optical dipole $n$P-exciton resonance transitions. These multiple peaks form asymmetric Lorentzian shapes, which are well formulated by Toyozawa \cite{Toyozawa1958}:
\begin{equation}\label{eq:toyozawa}
\alpha_{n\text{P}}(E) = C_{n\text{P}} \frac{\frac{\Gamma_{n\text{P}}}{2}+2 Q_{n\text{P}}\left(E-E_{n\text{P}}\right)}{\left(\frac{\Gamma_{n\text{P}}}{2}\right)^{2}+\left(E-E_{n\text{P}}\right)^{2}},
\end{equation}

\begin{figure}[htbp]
\centering
\includegraphics[width=1.0\columnwidth]{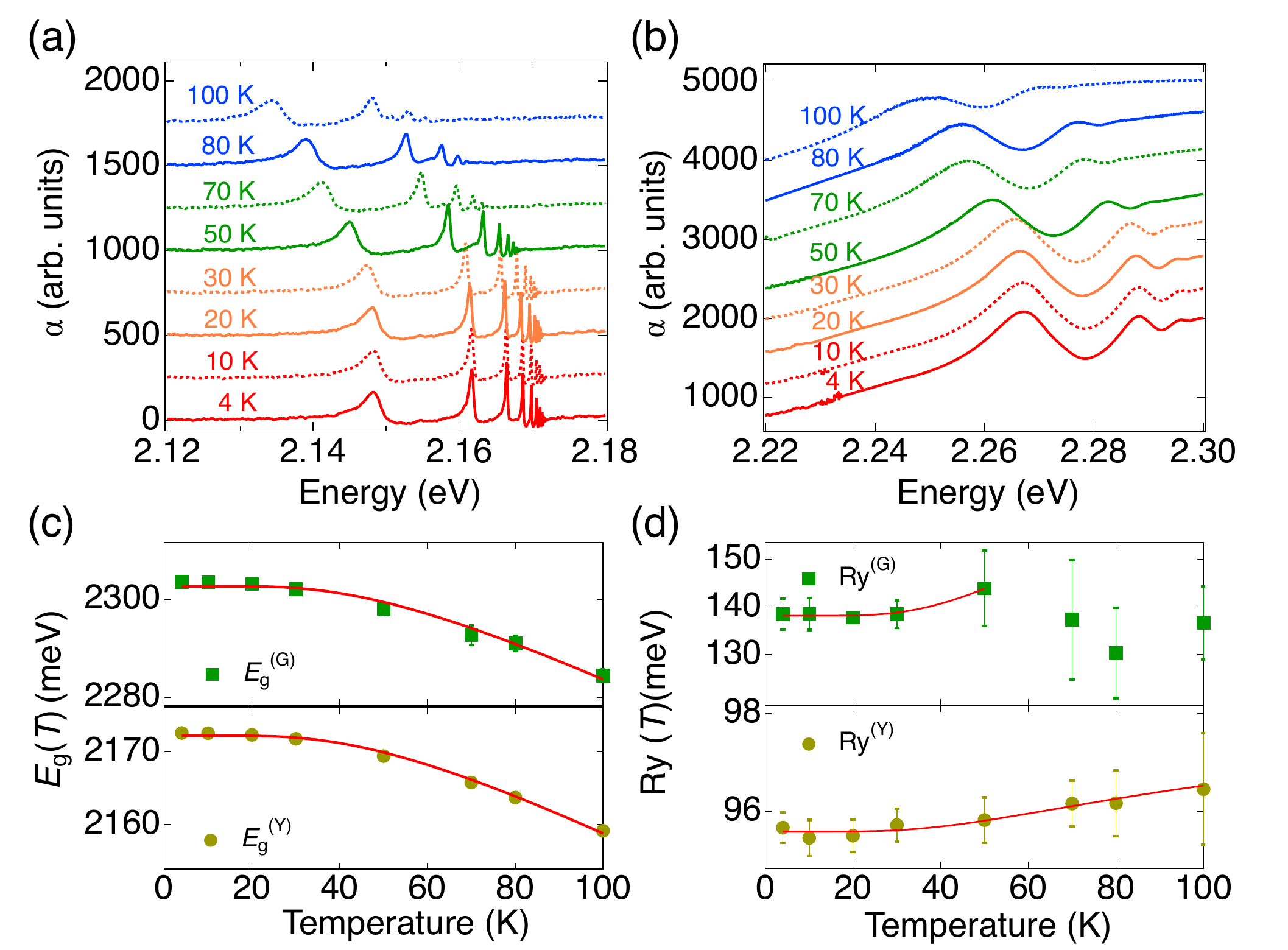}
\caption{$n$P exciton absorption spectra. (a) Yellow and (b) green $n$P-exciton resonance spectra taken at eight different temperatures after subtracting various background absorption contributions from bulk phonon-assisted absorption processes via ${\Gamma}_3^{-}$,  ${\Gamma}_4^{-}$ and minor phonons, as well as P-exciton non-resonant absorption through thermal broadening, and lastly the Urbach tail. For more details, refer to Appendix A. Each spectra is stacked up vertically by arbitrary offsets.
Temperature dependence of fitted band gap energy (c) and Rydberg binding energy. (d) for yellow (Bottom) and green (Top) P-excitons. Straight red lines are results from a fit to Eq. (\ref{eq4:Temp_RyEg}). Shades are standard deviations. The fitting parameters from Eq. (4) are in Table \ref{tab2:EgRyCotFitRes}.
}
\label{Fig2:nPExcitonAbsorption}
\end{figure}

\begin{table*}
  \caption{Coefficients of bandgap and Rydberg energies from the fitting with Eq. (\ref{eq4:Temp_RyEg}) for data in Figs. \ref{Fig2:nPExcitonAbsorption}(c) and \ref{Fig2:nPExcitonAbsorption}(d).}
    \label{tab2:EgRyCotFitRes}
\begin{ruledtabular}
  \begin{tabular}{c c c c c}  
        Color & $E_{\text{g0}}$ (meV) & $E_{\text{gT}}$ (meV) & $\text{Ry}_{\text{0}}$ (meV) & $\text{Ry}_{\text{T}}$ (meV) \\ 
        \hline
        Yellow & 2172.20  $\pm$ 0.18 & -25.93 $\pm$ 0.75 & 95.61 $\pm$ 0.05 & 1.76 $\pm$ 0.22\\  
        Green & 2302.70  $\pm$ 0.47  & -36.79 $\pm$ 2.03 & 139.07 $\pm$ 1.59 & -9.31 $\pm$ 6.84 \\  
    \end{tabular}
  \end{ruledtabular}
\end{table*}
\noindent where $C_{n\text{P}}$ is related to the oscillator strength that is computed from the absorption coefficient integrated over frequency, $\Gamma_{n\text{P}}$ is the linewidth of the exciton resonant peak at the energy $E_{n\text{P}}$, and $Q_{n\text{P}}$ represents the degree to which the Lorentzian lineshape is asymmetric.  
The isolated yellow and green $n$P-exciton resonance spectra at different temperatures are collected in Fig. \ref{Fig2:nPExcitonAbsorption}(a) and \ref{Fig2:nPExcitonAbsorption}(b), respectively. At 4 K, the highest $n$ = 11  of the yellow exciton resonant peak is clearly visible and stable, whereas the maximum $n$ of clear resonance peaks decreases down to $n$ = 5 at 100 K. Note that the maximum $n$ of the yellow exciton transitions observed in our low-temperature spectra (4 - 50 K) is limited by the spectral resolution of our monochromator ($\sim$80 $\mu$eV). The resolution-limited access to the highest Rydberg excitons can be overcome by the use of a narrow linewidth tunable laser. The upper bound of $n$-th observable Rydberg exciton states can be found by thermal energy at finite temperatures. The maximum $n$ = 5 Rydberg exciton state at 100 K is reasonably close to $\sqrt{E_B/k_BT} \sim 3.2$  where the binding energy $E_B$of the yellow excitons is 90 meV.  

On the other hand, the green $n$P-exciton resonance peaks are still on top of a residual continuous broad background signal even after the subtraction of the major phonon-assisted continuum and Urbach tail. This implies that green $n$P-excitons may be masked by other unknown absorption processes. We account for this remaining background spectrum by adding a constant offset term to our fitting. For more details on the background removal, see Appendix A. At all measurement temperatures, only $n$ = 2 - 4 green exciton peaks are observable, which was not due to our equipment resolution. Our result is consistent with other reports \cite{Itoh1975,Schone2017}.
We conjecture that the unobserved peaks with the higher numbered $n$ for green excitons below 50 K may be buried in the measurement noise and larger unknown phonon background absorption in our sample. This aspect needs to be further investigated in future.

The asymmetric Lorentzian-shaped peaks are fitted by Eq.(\ref{eq:toyozawa}) to extract various parameters: yellow and green exciton resonance energy $E_{n\text{P}}$, 
spectral full-width at half maximum (FWHM) linewidth $\Gamma_{n\text{P}}$, oscillator strength from $C_{n\text{P}}$, and asymmetry of the peaks $Q_{n\text{P}}$. In Appendix C, we explain the details of our Toyozawa-formula fitting procedure to capture $n$P-exciton peaks from spectra at the two extremes of our temperature measurement range (4 K and 100 K). Among these parameters, we first examine the yellow (Y) and green (G)-exciton band gap energy $E_{\text{g}}^{\text{(Y, G)}}$ and the excitonic Rydberg energy $\text{R}_{\text{y}}^{\text{(Y, G)}}$. According to the hydrogen-like Rydberg energy states,   the $n$P-exciton resonance energy $E_{n\text{P}}^{\text{(Y, G)}}$ is written similar to the famous Rydberg formula for hydrogen with $n$, $E_{\text{g}}^{\text{(Y, G)}}$, and $\text{R}_{\text{y}}^{\text{(Y, G)}}$ \cite{Schone2016QD},

\begin{equation}\label{eq:scalingEn}
E_{n \text{P}}^{\text{(Y, G)}} = E_{\text{g}}^{\text{(Y, G)}}-\frac{\text{Ry}^{\text{(Y, G)}}}{\left(n-\delta_{n}\right)^2},
\end{equation}
where the quantum defect $\delta_{n}$ accounts for non-hydrogenic features that may arise from the complex valence band structures \cite{Schone2016}. Since the actual values of $\text{Ry}_{\text{}}^{\text{(Y, G)}}$ and $\delta_{n}$ are closely connected, for simplicity, we neglect the quantum defect ($\delta_n = 0$) here and save the quantum defect analysis for Sec. III.
 
A straightforward fit of Eq.(\ref{eq:scalingEn}) with $\delta_n = 0$ gives $E_{\text{g}}^{\text{(Y, G)}}$ and $\text{Ry}_{\text{}}^{\text{(Y, G)}}$ for yellow and green series as a function of temperature plotted in Fig. \ref{Fig2:nPExcitonAbsorption}(c) and \ref{Fig2:nPExcitonAbsorption}(d), respectively. The red shift of $E_{\text{g}}^{\text{(Y, G)}}$ by temperature increase is clearly observed in both excitons, whose tendency agrees with the 1S-ortho exciton energy measured from PL experiments (See Fig. \ref{Fig1:PhononAssistedAbsorption}(c)). 
The temperature-dependence of these parameters can be explained by Eliott's theory similar to $E_{\text{1S}}(T) $ \cite{Itoh1975}:
\begin{equation}
\begin{array}{l}
E_\textrm{g}^{\text{(Y, G)}}(T)=E_\textrm{g0}^{\text{(Y, G)}} +
E_\textrm{gT}^{\text{(Y, G)}}\left( \operatorname{coth}\left(\frac{\hbar \omega_3}{2 k_B T}\right)-1\right),\\
\text{Ry}_\textrm{}^{\text{(Y, G)}}(T) = \text{Ry}_\textrm{0}^{\text{(Y, G)}} +
\text{Ry}_\textrm{T}^{\text{(Y, G)}}\left( \operatorname{coth}\left(\frac{\hbar \omega_3}{2 k_B T}\right)-1\right),
\end{array}\label{eq4:Temp_RyEg}
\end{equation}
where $E_\textrm{g0}^{\text{(Y, G)}}$ and $\text{Ry}_\textrm{0}^{\text{(Y, G)}}$ are temperature-independent terms. The terms of $E_\textrm{gT}^{\text{(Y, G)}}$ and $\text{Ry}_\textrm{T}^{\text{(Y, G)}}$ are associated with dominant $\Gamma_3^-$-phonons. Under the assumption of $n$-independence and $\hbar\omega_3$ = 13.6 meV, we obtain the fitting results of the Eq.~\ref{eq4:Temp_RyEg} parameter values, which are summarized in Table \ref{tab2:EgRyCotFitRes}. Our results are consistent with literature values taken at 4 K \cite{Itoh1975,Gross1956}. Theoretically, the values of $\text{Ry}^{\text{(Y)}}$ at 4 K are estimated by $\text{Ry}^{\text{(Y)}} = (\mu/\varepsilon^2) \text{Ry}$ with effective exciton mass $\mu$ and dielectric constant $\varepsilon$ = 7.5 from the hydrogen Rydberg constant Ry = 13.6 eV. Depending on the choice of the effective mass $\mu$ essentially from the band structure curvatures, $\text{Ry}^{\text{(Y)}}$ ranges between 86 and 96 meV ($\sim10 \%$ fluctuation) \cite{Zielinska-Raczynska2016}. Our value of $\sim$95.6 meV from hydrogen-like scaling laws falls within this range. Furthermore, Table \ref{tab3:RyQDFsurvey} provides the several Ry values of \ce{Cu2O} excitons chosen in the literature. 

Comparing the two exciton series, green $n$P-excitons have a temperature dependence term $E_{\textrm{gT}}$ which is larger than the yellow exciton term by about 42\%. Since both excitons are transiting from the same  $\Gamma_6^+$ conduction band, the larger $E_{\textrm{gT}}^{(\text{G})}$ indicates that the splitting between two valence-bands grows as temperature rises. Temperature dependence of the Rydberg energy is expected to originate from the lattice constant change at different temperatures, consequently modifying the band gap energy and the dielectric constant. The $\text{Ry}_\textrm{T}^{\text{(Y)}}$ exhibits a reasonable coincidence with that in Ref. \cite{Itoh1975}; however, the Ry values in green series are relatively insensitive to temperatures with large error bars probably due to the small number of accessible Rydberg states and broadness of the peaks, which may hinder the accurate evaluation of the temperature-effect.

\section{Scaling Laws}

This section is devoted to investigate the scaling laws of the yellow and green Rydberg exciton optical properties in terms of $n$ at each temperature between 4 K to 100 K based on the Toyozawa's asymmetric Lorentzian fitting analysis. Specifically, we examine four properties of $n$-dependence : the $n$P-exciton energies, the energy spacing between adjacent $n$P-exciton states, the lifetime of the states from the spectral linewidth, and oscillator strength proportional to the amplitude of absorption coefficient. We also look at the asymmetry of Lorentzian shapes. 

\subsection {Yellow Excitons}

\begin{figure*}[htbp]
\centering
\includegraphics[width=0.9\textwidth]{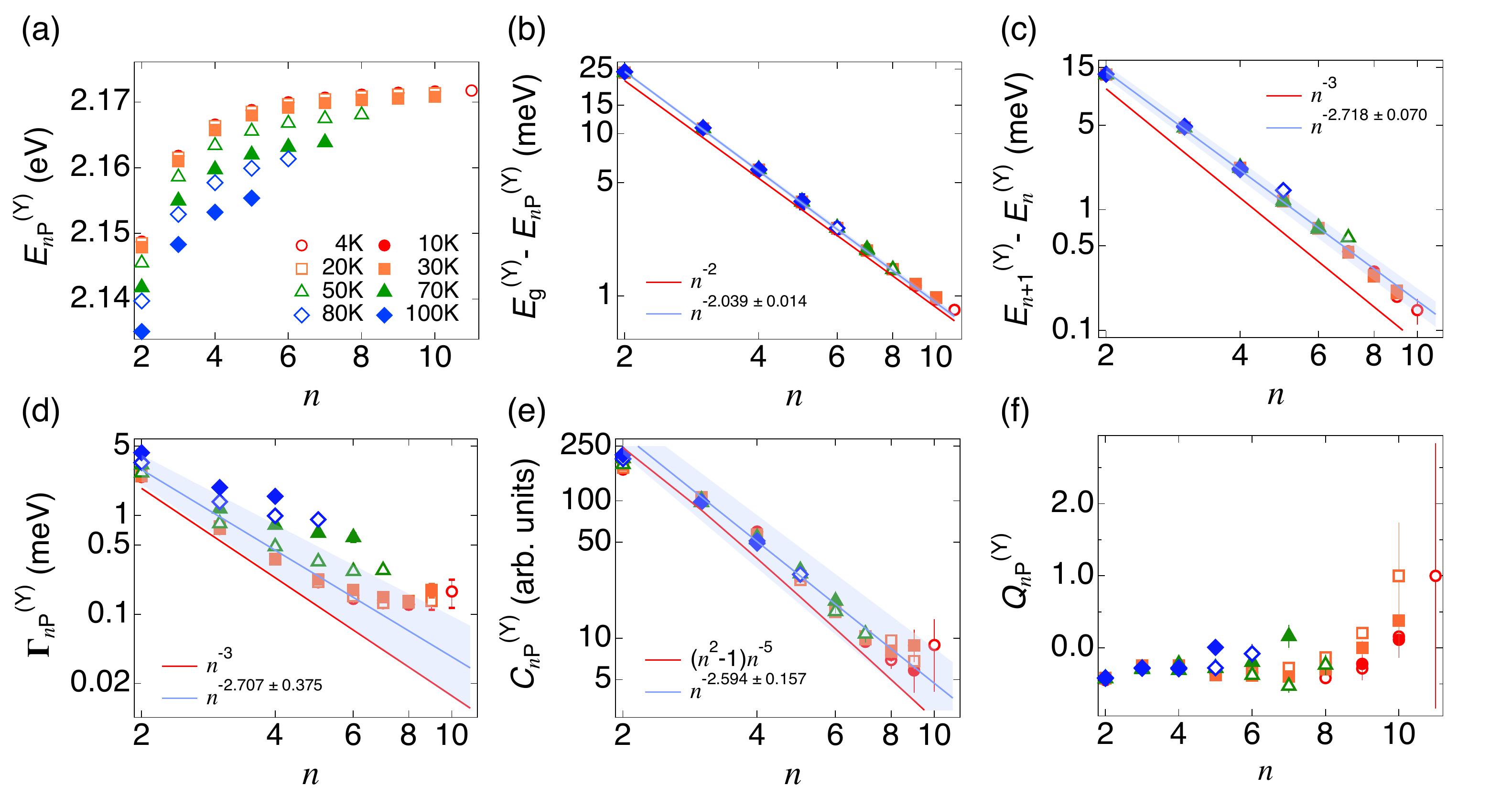}
\caption{Yellow P-exciton resonances at different temperatures:
(a) $E_{n\text{P}}^{\text{(Y)}}$, excitation resonance energy, (b) $E_{\text{g}}^{\text{(Y)}}-E_{n\text{P}}^{\text{(Y)}}$, exciton binding energy, 
(c) $E_{n+1}^{\text{(Y)}}-E_{n}^{\text{(Y)}}$
(d) $\Gamma_{n\text{P}}^{\text{(Y)}}$, the Full-width at half-maximum of exciton peaks,
(e) $C_{n\text{P}}^{\text{(Y)}}$, the peak areas of asymmetric Lorentzian peaks, and (f) the degree of asymmetry $Q_{n\text{P}}^{\text{(Y)}}$ of exciton resonance. Red straight lines in (b), (c), (d), and (e) represent the theoretical scaling behavior as guidelines, while the light blue lines with shades show the fitting results of the average exponent of $n$ with the statistical standard deviation.}
\label{Fig3:yellowScaling}
\end{figure*}

First, Rydberg-exciton resonance energies, $E_{n\text{P}}^{\text{(Y)}}$, are plotted against $n$ in a linear scale (Fig. \ref{Fig3:yellowScaling}(a)). At a given temperature, as $n$ is increased, $E_{n\text{P}}^{\text{(Y)}}$ gets saturated towards the bandgap energy $E_{\text{g}}^{\text{(Y)}}$, which follows well with Eq. \ref{eq:scalingEn}. The increase in temperature leads to the smaller values of $E_{n\text{P}}^{\text{(Y)}}$ (red-shift) closely related to the red-shift trend of $E_{\text{g}}(T)$ in Fig. \ref{Fig2:nPExcitonAbsorption} (c). In order to observe the scaling behavior, we remove this temperature dependence by computing $E_{\text{g}}^{\text{(Y)}} - E_{n\text{P}}^{\text{(Y)}}$ at all experimental temperatures. As can be seen in Fig. \ref{Fig3:yellowScaling}(b), this data is well fit by a single line, following the famous $n^{-2}$ scaling behavior. The exponent of $n$ from our linear regression analysis with all experimental plots in a log scale is on average -2.039 $\pm$ 0.014, with shading representing the uncertainty. Since we are able to detect the states of $n$ = 7 or 8 routinely above 50 K, we predict that the higher $n \sim$ 10 Rydberg states may be captured by high-resolution spectroscopy above liquid-nitrogen temperatures.  

The energy difference between the two-adjacent exciton energy $E_{\Delta n}$ is expected to exhibit a universal power law of $n^{-3}$ at each temperature from this simple derivation,

\begin{equation}\label{eq10:Edeltan}
\begin{aligned}
E_{\Delta n}^{\text{(Y)}} & = E_{n+1}^{\text{(Y)}}-E_{n}^{\text{(Y)}} =\frac{\text{Ry}^{\text{(Y)}}}{\left(n-\delta_{n}\right)^{2}}-\frac{\text{Ry}^{\text{(Y)}}}{\left((n+1)-\delta_{n+1}\right)^{2}}\\
&\cong \text{Ry}^{\text{(Y)}}\left(\frac{2 n+1+6 \delta}{n^{2}(n+1)^{2}}\right)
\approxprop n^{-3}.
\end{aligned}
\end{equation}
The evaluated $E_{\Delta n}^{\text{(Y)}}$ are included for yellow $n$P-excitons plotted in Fig. \ref{Fig3:yellowScaling}(c). Again the data for all temperatures converges into one line, and the  exponent value is -2.718 $\pm$ 0.070 (with the uncertainty represented by shading) and the theoretical guide line with $n^{-3}$ in red straight line (Fig. \ref{Fig3:yellowScaling}(c)). Our result matches well with the theory within a deviation of $\sim 10 \%$.

The next Rydberg property we examine is the FWHM, $\Gamma_{n\text{P}}^{\text{(Y)}}$, which is expected to have the dependence of  $n^{-3}$ \cite{Toyozawa1958}. Figure \ref{Fig3:yellowScaling}(d) collects experimental $\Gamma_{n\text{P}}^{\text{(Y)}}$, where we notice two different regimes. For low temperatures below 30 K, $\Gamma_{n\text{P}}^{\text{(Y)}}$ decreases along the $n^{-3}$ line below $n $ = 8 but pinned around 100  $\mu$eV above  $n $ = 8. This saturation results from the convolution of the exciton resonance and our monochromator spectral transfer function, whose minimum spectral resolution is $\sim$ 80 $\mu$eV. On the other hand, the increase of temperature over 30 K causes the departure from  the $n^{-3}$ trend for even lower $n $ states ($n<6$) owing to the thermal broadening. The average value of the power exponent is -2.707, which is not significantly deviated from -3, but the standard deviation is noticeably bigger due to the aforementioned reasons. Furthermore, we plot $C_{n\text{P}}^{\text{(Y)}}$ in Fig. \ref{Fig3:yellowScaling}(e) to assess the scaling of oscillator strength. The obtained $C_{n\text{P}}^{\text{(Y)}}$ reasonably match with the theoretical expectation of $(n^{2}-1)/n^{5}$ except the $n =$ 2 state \cite{Heckotter2017,Kruger2019}. All temperature data sit reasonably well-overlapped without any adjustment, which implies that the magnitude of the oscillator strength remains similarly in our temperature range. $C_{n\text{P}}^{\text{(Y)}}$ at the larger $n$-valued states is more sensitive to additional disturbance factors to yield the discrepancy from the scaling relation.
Most resonance peaks which fit very well to Eq. (\ref{eq:toyozawa}) exhibit strong asymmetry ($|Q_{n\text{P}}| > 0.3$), which reveals the interband effect associated with exciton band structures \cite{Toyozawa1958}. For example, the high energy tail in positive $Q_{n\text{P}}$ implies the Fano-type exciton-continuum part interference.  At all temperatures, $Q_{n\text{P}}$ starts with negative values in smaller $n$ numbers but progressively moves to the positive side. For $n \geq 10$ we see a strong trend towards positive asymmetry with a high degree of uncertainty, which we attribute to the resolution limit of our spectrometer.

\begin{table*}[htbp]
   \caption{Ry and $\delta$ values to evaluate yellow P-exciton resonance energies in \ce{Cu2O}. $^\dagger$Phys. Rev. B 23, 2731 (1981).  $\ast$Chosen values for fitting to experimental data in Ref.~\cite{Kazimierczuk2014}.}
    \label{tab3:RyQDFsurvey}
      \begin{ruledtabular}
  
    \begin{tabular}{cccc}
        Reference & Ry (meV) & $\delta$ & Temperature (K) \\  
        \hline
        Ziemkiewicz {\it{et al.}} (2020)\textsuperscript{\cite{Ziemkiewicz2020}} & 91.5$^\ast$ & 0.083$^\ast$ & 10 \\
        Stolz {\it{et al.}}(2018)\textsuperscript{\cite{Stolz2018}} & 87.653$^\ast$ & 0.1986$^\ast$ & 1.2  \\ 
        Sch\"one {\it{et al.}} (2017) \textsuperscript{\cite{Schone2017}} & 87 &  0 & 2  \\
        Schöne {\it{et al.}} (2016)\textsuperscript{\cite{Schone2016QD}} & 86.04 & $\sim$0.32 ($n > 10)$ &   \\ 
        Kazimierczuk {\it{et al.}} (2014)\textsuperscript{\cite{Kazimierczuk2014}} & 92  & 0.23 & 1.2 \\  
        Uihlein {\it{et al.}} (1981)\textsuperscript{\cite{Uihlein1981}}& 87 & 0  & 4.2 \\ 
        Itoh {\it{et al.}} (1975)\textsuperscript{\cite{Itoh1975}} & 93 & 0 &  \\ 
         \end{tabular}
    \end{ruledtabular}
\end{table*}

\begin{figure}[htbp]
\centering
\includegraphics[width=1\columnwidth]{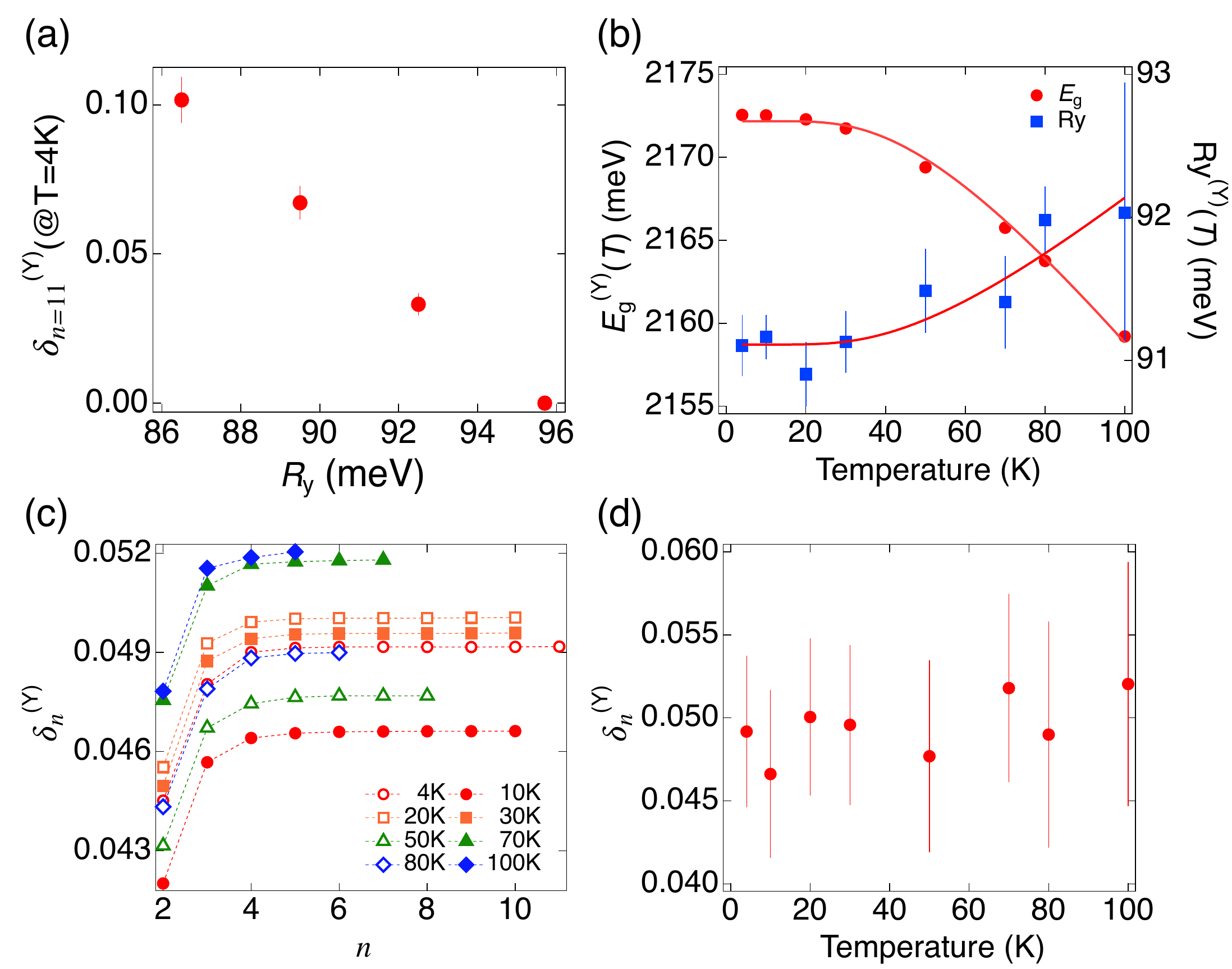}
\caption{Yellow-exciton quantum defect $\delta_n^{(\text{Y})}$(T) study results as a function of $n$ and temperature. (a) An example of $\delta$ fitting results when $n$=11 at 4K with different Ry. Decreasing $\delta$ trend is observed as the Ry increases. (b) Fitted yellow-exciton $E_g^{(\text{Y})}$ and $\text{Ry}^{(\text{Y})}$ considering quantum defect with Ry $\sim$ 91 - 92 meV which is obtained from $\delta_n^{(\text{Y})}$(T) study. Straight lines are the fits to Eq. (\ref{eq4:Temp_RyEg}). (c) $\delta_n^{(\text{Y})}$ vs $n$ with $E_g^{(\text{Y})}$ and Ry which are presented in Fig. \ref{eq4:Temp_RyEg}(b). (d) The temperature dependent $\delta_n^{(\text{Y})}$. The red points are average $\delta_n^{(\text{Y})}$ and red lines are error bar.}
\label{Fig4:qdf}
\end{figure}

\begin{table}[h]
  \caption{Coefficients of bandgap and Rydberg energies from the fitting with Eq. (\ref{eq4:Temp_RyEg}) for data in Fig. \ref{Fig4:qdf}(b).}
    \label{tab4:EgRyCotFitRes_QDF}
        \begin{ruledtabular}
 
    \begin{tabular}{c c c c} 
        $E_{\text{g0}}$ (meV) & $E_{\text{gT}}$ (meV) & $\text{Ry}_{\text{0}}$ (meV) & $\text{Ry}_{\text{T}}$ (meV) \\ 
        \hline
     
        2172.20  $\pm$ 0.18 & -25.77 $\pm$ 0.15 & 91.11 $\pm$ 0.08 & 1.99 $\pm$ 0.34\\  
       
    \end{tabular}
    \end{ruledtabular}
 \end{table}

Our analysis so far neglects the effect of the quantum defect by setting $\delta_n$ = 0 in Eq.~(\ref{eq:scalingEn}); however, the non-hydrogenic nature was reported in \ce{Cu2O} excitons \cite{Schone2016}. Now we discuss the effect of quantum defect in our temperature-dependent data and evaluate $E_\text{g}$ and Ry with non-zero $\delta_n$ according to Eq.~(\ref{eq:scalingEn}). At a fixed temperature, the bandgap energy $E_\text{g}$ values remain almost the same between zero and non-zero $\delta_n$, with a difference smaller than 0.002\%, whereas the value of Ry is extremely sensitive to the value of $\delta$ in $\delta_n^{(\text{Y})}$ study in Fig. \ref{eq4:Temp_RyEg}(a). The plot shows $\delta_n^{(\text{Y})}$ as a function of different Ry values at 4 K. As can be seen, $\delta_n^{(\text{Y})}$ decreases as Ry increases. Table \ref{tab3:RyQDFsurvey} surveys the Ry and $\delta$ values in a small set of the literature. 
Ry values sit between 86 and 93 meV with a wide range of the $\delta_n$ values between 0 and around 0.32. Equation (\ref{eq:scalingEn}) tells us that Ry and $\delta_n$ are anticorrelated for a fixed $E_{n\text{P}}$, namely a smaller Ry accompanies a larger $\delta_n$. Indeed, this trend appears in the Table \ref{tab3:RyQDFsurvey}; however, a concerning point is that even for the same experimental Rydberg data \cite{Kazimierczuk2014}, the chosen $\delta$ are spread out significantly while Ry values are within 10 - 11\% range similar to the theoretical Ry uncertainty $\sim$10 $\%$ \cite{Kazimierczuk2014,Ziemkiewicz2020,Schone2016}. Note that extensive theoretical discussions are given about central-cell corrections in Rydberg excitons in {\ce{Cu2O}}, where several factors are explicitly treated, including the nonparabolicity of the bands which can modify the bare electron and hole masses, the momentum and frequency dependent dielectric function associated with the coupling of electrons and holes to the longitudinal optical phonons, and exchange interactions~\cite{Kavoulakis1997, Schone2016QD, Schone2016, Uihlein1981}. Thus, we remark that the non-converging values of $\delta$ and Ry from experimental data would require further careful studies both in experiments and theoretical calculations.

Despite this uncertainty, we show our best attempt to assess our experimental data by incorporating the quantum defect $\delta$. We assume that both $E_\text{g}$ and Ry are independent of $n$ at a given temperature. Inspired by previous work \cite{Schone2016QD}, where $\delta_n$ approaches steady values above $n\geq 4-5$, we evaluate $\delta_n$ vs. $n$  for the  $n\geq 4-5$ data. Figure \ref{Fig4:qdf}(b) shows the updated $E_\text{g}$ and Ry using this method with non-zero $\delta_n$ for the yellow excitons. Ry is taken by the $\delta_n^{(\text{Y})}$ study, finding optimized Ry values through  the standard deviation of $\delta_n^{(\text{Y})}$ with different $\delta_n^{(\text{Y})}$ values. Both $E_\text{g}$ and Ry still obey hyperbolic cotangent temperature relations in Eq. (\ref{eq4:Temp_RyEg}). The fitting results of two cases are summarized in Table \ref{tab4:EgRyCotFitRes_QDF}. Compared to Figs. \ref{Fig2:nPExcitonAbsorption}(c) and (d), $E_\text{g}$ is almost the same, within a 2\% difference, while Ry with the non-zero $\delta_n$ is between 91 and 92 meV, about 5\% lower than earlier Ry $\sim$ 95 - 96 meV. This discrepancy agrees with the previous reports which consider Ry with or without quantum defects \cite{Gross1956, Schone2016, Heckotter2017}. Regarding Fig. {\ref{Fig4:qdf}(d)}, $\delta_n$ appears to have no observable trend over the temperature range considered here, with the largest deviation from the mean of the values being only 11\%. Thus, we assume $\delta_n$ to be constant as a function of temperature.

\subsection {Green Excitons}

\begin{figure*}[htbp]
\centering
\includegraphics[width=0.95\textwidth]{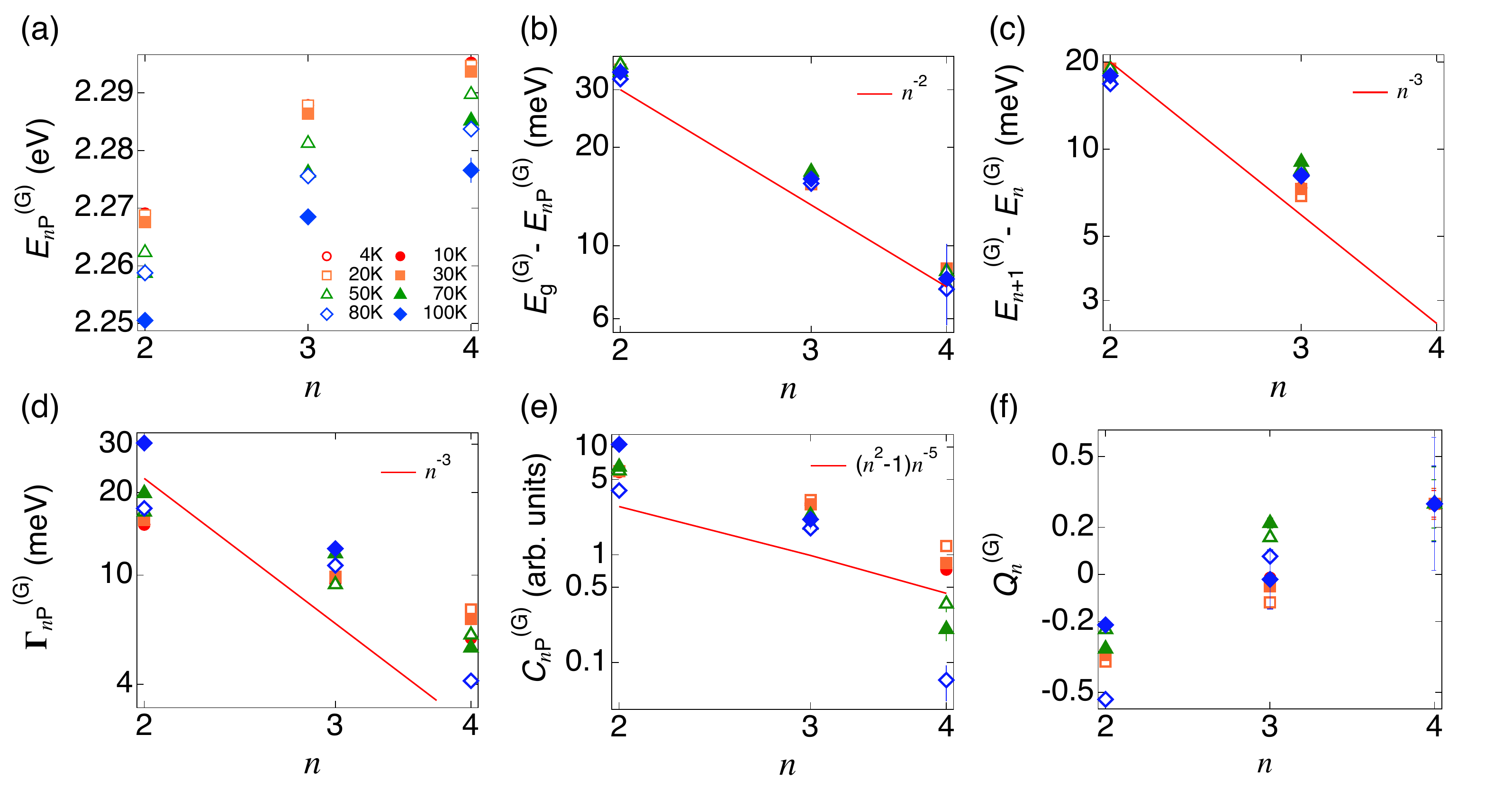}
\caption{Green P-exciton resonances at different temperatures:
(a) $E_{n\text{P}}^{\text{(G)}}$, excitation resonance energy, (b) $E_{\text{g}}^{\text{(G)}}-E_{n\text{P}}^{\text{(G)}}$, exciton binding energy, (c) $E_{n+1}^{\text{(G)}}-E_{n}^{\text{(G)}}$,
(d) $\Gamma_{n\text{P}}^{\text{(G)}}$, the Full-width at half-maximum of exciton peaks,
(e) $C_{n\text{P}}^{\text{(G)}}$, the peak areas of asymmetric Lorentzian peaks, and (f) the degree of asymmetry $Q_{n\text{P}}^{\text{(G)}}$ of exciton resonance. Red straight lines in (b), (c), (d), and (e) represent the theoretical scaling behavior as guidelines.}
\label{Fig5:greenScaling}
\end{figure*}

For green $n$P-excitons, we can distinguish only three resonance peaks ($n$ = 2, 3, 4) at all temperatures so that we encounter technical challenges in performing reliable scaling analysis. Despite the existence of increasing residual background absorption, we apply the same hydrogen-like exciton model to green absorption data (Fig. \ref{Fig2:nPExcitonAbsorption}(b)), similarly to yellow excitons. The $n$-dependence of various parameters at different temperatures is presented in Fig. \ref{Fig5:greenScaling}. 

Using the green band-gap and Rydberg energies reported in Figs. \ref{Fig2:nPExcitonAbsorption}(c) and (d),  we are able to pinpoint  $E_g^{\text{(G)}} - E_{n\text{P}}^{\text{(G)}}$ of the green $n$P-excitons, which merge into a single line (Fig. \ref{Fig5:greenScaling}(b)). We find that $E_g^{\text{(G)}} - E_{n\text{P}}^{\text{(G)}}$ has an excellent agreement with $n^{-2}$.
The energy gap between adjacent states in Fig. \ref{Fig5:greenScaling}(c) reduces as $n$ goes up along the $n^{-2}$-trend, but two data points cannot validate our linear regression analysis result conclusively. 

We notice the different spectral linewidth trend of green exciton from that of yellow excitons.  $\Gamma_{n\text{P}}^{\text{(G)}}$ is much broader than that of the yellow excitons. 
Although $\Gamma_{n\text{P}}^{\text{(G)}}$ (Fig.~\ref{Fig5:greenScaling}(d)) of the green exciton resonance is broader than our minimum spectral resolution, it still does not follow the expected scaling law of $n^{-3}$ like the yellow excitons. We may understand the linewidth of the green excitons in terms of several factors different from the yellow excitons. Similar to the yellow excitons, the green excitons also couple to various components of $\Gamma_3^-, \Gamma_4^-$ phonons and non-resonant Urbach tail. Since the green exciton peaks are located in the yellow absorption continuum region, they are coupled to free electron-hole pairs in the yellow continuum via the valence bands so that they are not truly bound like the yellow excitons. The valence bands $^4\Gamma_8^+$ are degenerate with two different effective masses, which may play a role in the broader spectral linewidth of the green excitons. Recently, Rommel \textit{et al.} calculate that the linewidths from the yellow-continuum coupling only follow the $n^{-3}$ scaling law \cite{Rommel2020}. Thus, we conjecture that if we can isolate this yellow-continuum contribution only, we may observe the expected scaling law of the spectral linewidths in the green excitons.

Significant thermal broadening is also seen at temperatures $\geq$ 50 K, indicating stronger phonon scattering. 
The $C_{n\text{P}}^{\text{(G)}}$ of the green $n$P-excitons are 20 - 30 times larger than those of their yellow counterparts. The theoretically derived ratio of oscillator strength between yellow and green excitons is 15 \cite{Elliott1957,Itoh1975}, which is on the same order as our estimate.
From the trend of $C_{n\text{P}}^{\text{(G)}}$ in  Fig. \ref{Fig5:greenScaling}(e), we guess that the oscillator strength will decrease much more rapidly than the scaling law of $(n^{2}-1)/n^{5}$. The origin of this departure from the known scaling law may be common to the cause of the broadened $\Gamma_{n\text{P}}^{\text{(G)}}$. Some non-radiative processes in green excitons can broaden the resonance linewidth and reduce the radiative path, and consequently the oscillator strength.  
The green exciton resonance peaks are also asymmetric Lorentzians with non-zero negative $Q_{n\text{P}}^{\text{(G)}}$ at $n=2$, which moves to positive $Q_{n\text{P}}^{\text{(G)}}$ at $n=4$. This transition resembles the yellow-exciton $Q_{n\text{P}}^{\text{(Y)}}$. Unlike the yellow exciton series, we are unable to perform  the reliable quantum defect analysis for the green excitons due to only having 2-3 data points. 

\section{Temperature dependence of Rydberg Exciton properties: spectral linewidth, lifetime, peak areas, and asymmetry}

In this section, we focus on the temperature behavior of the experimental parameters.  The temperature dependence of the yellow and green exciton resonant spectra has been studied previously in several works \cite{Itoh1975,Iwamitsu2012}; however, these discussions were mainly restricted to the 2P-exciton only. We expand the temperature study to higher $n$P-excitons for the purpose of high-temperature solid-state Rydberg applications. The assessment of the temperature-independent and dependent components using Eqs. (\ref{eq:Gamma}) and (\ref{eq:cnp}) provides the intuitive view for how the phonon-originated broadening is critical in the measured spectra at higher temperatures.

We revisit the temperature-dependent 1S-ortho exciton energy, bandgap energy, and Rydberg binding energy obtained by fitting the hyperbolic cotangent functions explicitly written in Eqs. (\ref{eq:Tdep}) and (\ref{eq4:Temp_RyEg}). The average value of $E_{\text{1S}} (0)$ at the zero-temperature limit is 2.0615 eV, and the $E_{\text{1S}} $ at $T$ = 4 K of 2.0332 eV almost perfectly matches with the reported value of 2.0327 eV within 0.03 \% error \cite{Biccari2009}. 
The coefficient $c_{\text{1S}}$ of the hyperbolic cotangent function in Eq. (\ref{eq:Tdep}) is expressed as $S \hbar \omega_3$, where $S = 1.89$ is related to the material constant and $\hbar \omega_3$ = 13.6 meV \cite{Biccari2009}. The values of $c_{\text{1S}}$ in Table \ref{tab1:E1SCotFitRes} give $S$ = 2.128 $\pm$ 0.071, around 13 \% higher than the expected value of 1.89. Similarly, the zero-temperature $E_{g0}$ and Ry$_0$ values are consistent with the literature values \cite{Itoh1975}. In this case, $E_{gT}$ yields $S \sim$ 1.906 with $\hbar \omega_3$ = 13.6 meV, within 1 \% error of the expected value of 1.89.

\begin{figure*}[t]
\centering
\includegraphics[width=\textwidth]{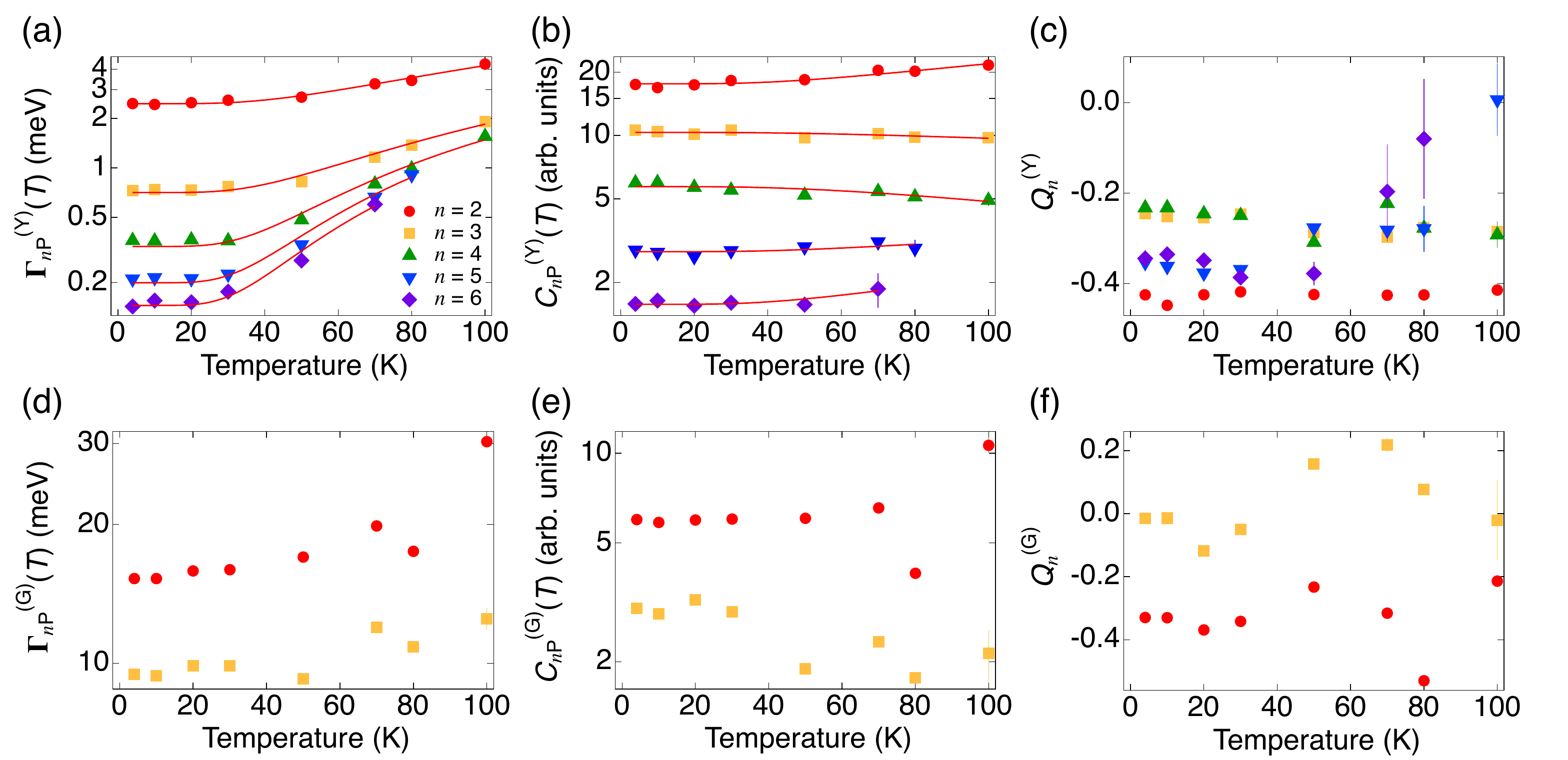}
\caption{Temperature-dependence of optical properties: the full-width at half maximum linewidths ($\Gamma_n$), amplitude of the exciton resonance ($C_n$), and asymmetry ($Q_n$) of yellow ((a),(b),(c))  and green ((d),(e),(f)) excitons.}
\label{Fig6:TemperatureDependence}
\end{figure*}

\begin{table*}[htp]
  \caption{Summary of fitted coefficient values of Eq. \ref{eq:Gamma} and \ref{eq:cnp} to the $\Gamma_n$ and $C_n$ data for yellow excitons in Fig. \ref{Fig6:TemperatureDependence}.}
    \label{tab5:CnGamFittingRes}
    \centering
      \begin{ruledtabular}
    \begin{tabular}{c c c c c } 
            $n$ & $\Gamma_{n\text{P} ,0}^{\text{(Y)}}$ (meV) & $\Gamma_{n\text{P}, T}^{\text{(Y)}}$ (meV) & $C_{n\text{P} ,0}^{\text{(Y)}}$ (arb. units) & $C_{n\text{P}, T}^{\text{(Y)}}$ (arb. units) \\  
        \hline
   
        2 & 2.46  $\pm$ 0.05 & -3.35 $\pm$ 1.26 & 0.1760 $\pm$ 0.00306 & -0.0835 $\pm$ 0.0831 \\  
        3 & 0.71 $\pm$ 0.03 & -2.19 $\pm$ 0.86 & 0.1030 $\pm$ 0.00159 & 0.0124 $\pm$ 0.0430 \\ 
        4 & 0.33 $\pm$ 0.03 & -2.22 $\pm$ 0.81 & 0.0572 $\pm$ 0.00131  & 0.0165 $\pm$ 0.0356  \\ 
        5 & 0.12 $\pm$ 0.13 & -3.45 $\pm$ 3.63 & 0.0258 $\pm$ 0.00389 & -0.0463 $\pm$ 0.1060  \\ 
        6 & 0.12 $\pm$ 0.07 & -2.83 $\pm$ 3.72 & 0.0148 $\pm$ 0.00268 &  -0.0462 $\pm$ 0.1370 \\ 
          \end{tabular}
       \end{ruledtabular}
\end{table*}

We observe the thermal broadening of the yellow- and green-exciton spectra lines in Figs. \ref{Fig6:TemperatureDependence}(a) and \ref{Fig6:TemperatureDependence}(e): the higher temperatures, the broader the exciton resonance peaks arising from phonons. Specifically, $\Gamma_n$ of the exciton resonance increases more rapidly above 60 K, and the degree of broadening is $n$ dependent. The thermal impact is more noticeable for higher $n$P-yellow excitons. This phonon-related broadening is formulated as a function of temperature with a dominant $\Gamma_3^-$-phonon process via \cite{Itoh1975},
\begin{equation}\label{eq:Gamma}
\Gamma_{n\text{P}}^{\text{(Y,G)}}(T) 
= \Gamma_{n\text{P} ,0}^{\text{(Y,G)}}+\Gamma_{n\text{P}, T}^{\text{(Y,G)}}\left( \operatorname{coth}\left(\frac{\hbar \omega_3}{2 k_B T}\right)-1\right).
\end{equation}
The fitting results of the first five yellow excitons are summarized in Table \ref{tab5:CnGamFittingRes} and plotted in Figs. \ref{Fig6:TemperatureDependence}(a) and (b). A linear relation between $\Gamma_{n\text{P}}^{\text{(Y)}}(T)$ and \textit{T} for $n$P excitons ($n$ = 2 - 6) is observed above 50 K. This observation is explained from the first two terms in the Taylor expansion of the hyperbolic cotangent function. From the theoretical asymptotic approximation, the linear term in $T$ is dominant above 45.6 K, which agrees well with our measurements and fitting results. The temperature effect on the yellow $n$P-exciton linewidth is primarily influenced by the $\Gamma_3^-$-phonon process, which is consistent with the earlier study of the yellow 2P-exciton \cite{Itoh1975}. 

The exciton lifetimes $\tau_n$ are estimated from the FWHM $\Gamma_n$ through $\tau_{n} \approx h/2\Gamma_n$, assuming lifetime is the dominant contribution to the linewidth {\cite{Kazimierczuk2014}}. For example, the linewidth of a yellow exciton with $n=7$, about 0.2 meV, corresponds to the lifetime of $\sim$ 20 ps as a lower bound.  At higher temperatures, the $n$P yellow excitons are short-lived and the impact of the temperature is significant in higher $n$P excitons. 
Since the yellow exciton peaks are much narrower than the green ones, yellow exciton lifetimes are longer by a factor of 10 to 50 times. The weaker temperature-sensitivity of green exciton lifetimes is the consequence of the much broader linewidths. 

The $C_{n\text{P}}^{\text{(Y,G)}}(T)$ vs. $T$ at different $n$ are drawn in Figs. \ref{Fig6:TemperatureDependence}(b) and \ref{Fig6:TemperatureDependence}(e). In the same manner as the linewidth, we made an attempt to express the temperature dependence of oscillator strength term $C_{n\text{P}}^{\text{(Y,G)}}(T)$ with the $\operatorname{coth}$-function given by 
\begin{equation}\label{eq:cnp}
C_{n\text{P}}^{\text{(Y,G)}}(T) = C_{n \text{P}, 0}^{\text{(Y,G)}} + C_{n \text{P}, T}^{\text{(Y,G)}}\left( \operatorname{coth}\left(\frac{\hbar \omega_3}{2 k_B T}\right)-1\right).
\end{equation}
The fitting coefficients for the $C_{n\text{P}}^{\text{(Y,G)}}(T)$ vs. $T$ for different $n$ are displayed in Table \ref{tab5:CnGamFittingRes}. 
Lastly, the temperature behavior of yellow and green exciton  $Q_{n\text{P}}^{\text{(Y,G)}}(T)$ are presented in Figs. \ref{Fig6:TemperatureDependence}(c) and \ref{Fig6:TemperatureDependence}(f). When we examine the yellow-exciton $Q_{n\text{P}}^{\text{(Y)}}(T)$-$T$ data, we observe that low-lying $n$ states preserve their asymmetric Lorentzian shapes until 100 K; however, higher $n$ states become symmetric above 50 K related to the thermal broadening. We omit the $n$ = 4 case for the green exciton in Figs. \ref{Fig6:TemperatureDependence}(d)-(f) since the results in these cases are inconclusive and off-trend.

\begin{table*}[htbp]
     \caption{Comparison between the properties of Rydberg excitons at $n=10$ and $n=25$ to Rydberg atoms at $n=50$ and $n=80$. $C_3$ is dipole-dipole interaction strength.\\ 
    $^*$Radiative lifetime is estimated through $h/2\Gamma_n$ from the spectral linewidth $\Gamma_n$ reported in Ref.~\cite{Kazimierczuk2014}. \\
    $^{**}$The van der Waals coefficients $C_6$ for Rydberg excitons are assumed to be $n$P-$n$P asymptote for $|M|=0$, the first row in the Table 1 of Ref.{~\cite{Walther2018}}.\\  
    $^{\mathsection}$The Blockade radius of Rydberg excitons is estimated from the equation $\sqrt[6]{C_6/(\Gamma_n/2)}$.\\
    $^\dagger$The data are obtained from the Alkali Rydberg Calculator. The radiative lifetime are for 1S$_{1/2}$, 60P$_{1/2}$ and 80P$_{1/2}$. The dipole-dipole interaction coefficient $C_3$ is between the states $|$60P$_{3/2}$,60P$_{3/2} \rangle$ and $|$61S$_{1/2}$,60S$_{1/2} \rangle$ or $|$80P$_{3/2}$,80P$_{3/2} \rangle$ and $|$81S$_{1/2}$, 80S$_{1/2} \rangle$.
    The van der Waals interaction coefficient $C_6$ is from the second-order perturbation theory. \\
    $^\ddagger$The atomic Rydberg blockade radius is calculated for the laser linewidth 3.0 MHz.}
    \centering
     \begin{ruledtabular}
   
    \begin{tabular}{c c c c c c c c} 
         & \multicolumn{3}{c}{Yellow, \ce{Cu2O}} & \multicolumn{3}{c}{Cs} &   \\ 
         quantity & $n$ = 2 & $n$ = 10  & $n$ = 25 & $n$ = 6 & $n$ = 60 &  $n$ = 80 & Unit \\ 
        \hline
         Binding energy  (Ry/$n^2$) & 0.023 & 0.009 & 0.0015 & 3.8 & 0.008 & 0.0017 & eV \\  
         Orbital/atomic radius ($\langle r \rangle $) & 4 & 164 & 1030 & 0.340 & 503 & 980 & nm \\       
         Radiative lifetime & 0.0002$^*$ &  0.026   &  1.03  &  30$^\dagger$ & 5.60e5  & 1.4e6 & ns \\ 
         Dipole-Dipole interaction ($C_3) $ &  & 6$^*$ & 234$^*$ &   &  54$^\dagger$  & 343$^\dagger$  & $\mu$eV$\cdot \mu$m$^3$ \\   
         van der Waals interaction ($|C_6|$) &  & 4.6$^{**}$ &  1.49e5$^{**}$  &   & 1.2e3$^\dagger$  & 1.33e5$^\dagger$ & $\mu$eV$\cdot \mu$m$^6$ \\   
         Blockade radius  &  & 0.7$^{\mathsection}$  & 7.3$^{\mathsection}$ &  & 6$^\ddagger$ & 11$^\ddagger$ & $\mu$m \\  
     \end{tabular}
     \end{ruledtabular}
     \label{tab6:RydAtomXton}
\end{table*}

\section{Discussions and Conclusion}

We have provided a thorough analysis of the yellow and green P-exciton absorption spectra in \ce{Cu2O}, focusing on the behavior of various scaling laws at different temperatures, specifically, in terms of $E_n$, $E_g-E_n$, $E_{\Delta n}$, the FWHM, oscillator strength, and asymmetry for the Lorentzian excitonic resonance peaks. Up to temperatures of 100 K, for $n$ not limited by the spectral resolution of our monochromator, yellow P-excitons exhibit the expected scaling behaviors, whereas only the first three green excitons are seen.

Our observation of Rydberg scaling behavior in \ce{Cu2O} excitons is encouraging for the development of a Rydberg solid-state system for both fundamental and applied research and engineering applications. In comparison to Rydberg atoms, the evaluation of Rydberg excitons in semiconductors facilitates an understanding of Rydberg physics in the presence of large number of electrons by identifying similarities and differences between atomic and solid-state systems. Furthermore, quantum Rydberg-exciton technologies may offer advantageous features such as compact footprints, fast operations, and integrated architectures via advanced semiconductor lithographic and etching processes.  Other advantages include that the excitons are ``stationary" in a host material, which can be in principle be created on demand on a particular position space. For Rydberg atoms, however, the atoms would need to be rearranged to the particular configurational state. 

One notable benefit in Rydberg excitons is that even the $n$ = 10 state may still enjoy reasonably strong Rydberg properties because of smaller Rydberg energy, larger Bohr radius, stronger interaction term and bigger Blockade radius. In Table \ref{tab6:RydAtomXton}, we compare our \ce{Cu2O} yellow $n=10$ P-excition with Caesium (Cs) Rydberg atom in terms of several quantities. For example, large electric dipole moment $e \langle r \rangle$ with the elementary charge  $e$ is beneficial when one Rydberg particle interacts with another. Table~\ref{tab6:RydAtomXton} reads that the $n$ = 25 yellow P-exciton of \ce{Cu2O} would possess the characteristic dipole matrix element as large as that of $n$ = 80 Cs Rydberg atom. Therefore, we learn that the dipole-dipole interaction $C_3$ strength, the van der Waals interaction $|C_6|$ strength and the blockade radius are comparable in both systems.
   
The study of solid-state Rydberg quantum platforms, while an interesting problem, still remains in its infancy. In particular, most studies of Rydberg excitons take place at low temperatures. Motivated by the numerous research activities to search for RT solid-state quantum systems, in this work, we present our spectroscopic measurements and analysis of Rydberg excitons from 4 K to 100 K in order to assess the feasibility of higher temperature Rydberg applications using \ce{Cu2O}. Although the number of observed resonance peaks in simple absorption spectra decreases at high temperatures, the Rydberg exciton states are robust and continue to exhibit the expected scaling behavior in the Rydberg exciton energies, spectral linewidth, and oscillator strength, even over liquid nitrogen temperatures. Our studies tell us that RT Rydberg excitons may not be advantageous in \ce{Cu2O} since the stable maximum $n$ would be 3 or 4 limited by thermal energy. However, we think high-temperature Rydberg systems would surely be an interesting venue for thorough studies at elevated temperatures. In contrast to viable solid-state quantum defects which actually still require cryogenic operations for sharp optical transition lines and also need a mediating bus for a submicron-ranged particle-particle interaction, Rydberg excitons inherently enjoy direct dipole-dipole and/or van der Waals interactions. Their sufficiently longer lifetime would bring a plus compared to the interaction coefficients. Hence, we believe that our temperature studies provide promising results to access robust Rydberg exciton states with reasonably high $n$ values around 10 for yellow series above 70 K. Future research plans may include quantifying the Rydberg blockade~\cite{Khazali2017,Heckoetter2020}, defining a qubit and demonstrating single and two-qubit operations via standard and sophisticated optical techniques at high temperatures.

\bibliography{Cu2Obib}

\section*{Acknowledgments}
D.D.K., A.G., H.B.Y., and N.Y.K. acknowledge the support of Industry Canada, the Ontario Ministry of Research \& Innovation through Early Researcher Awards, the Canada First Research Excellence Fund-Transformative Quantum Technologies (CFREF-TQT), and Canada Foundation for Innovation 2017 Innovation Fund program (CFI-IF Project 36255). D.D.K. thanks T. Yoon for his help to build an optical setup. N.Y.K. thanks Prof. Makoto Kuwata-Gonokami for sample preparation support and fruitful discussions.
  
\clearpage
\section*{Appendix}

\subsection*{A. Estimation of the environmental background absorption}

\begin{figure}[htbp]
\centering
\includegraphics[width=1.0\columnwidth]{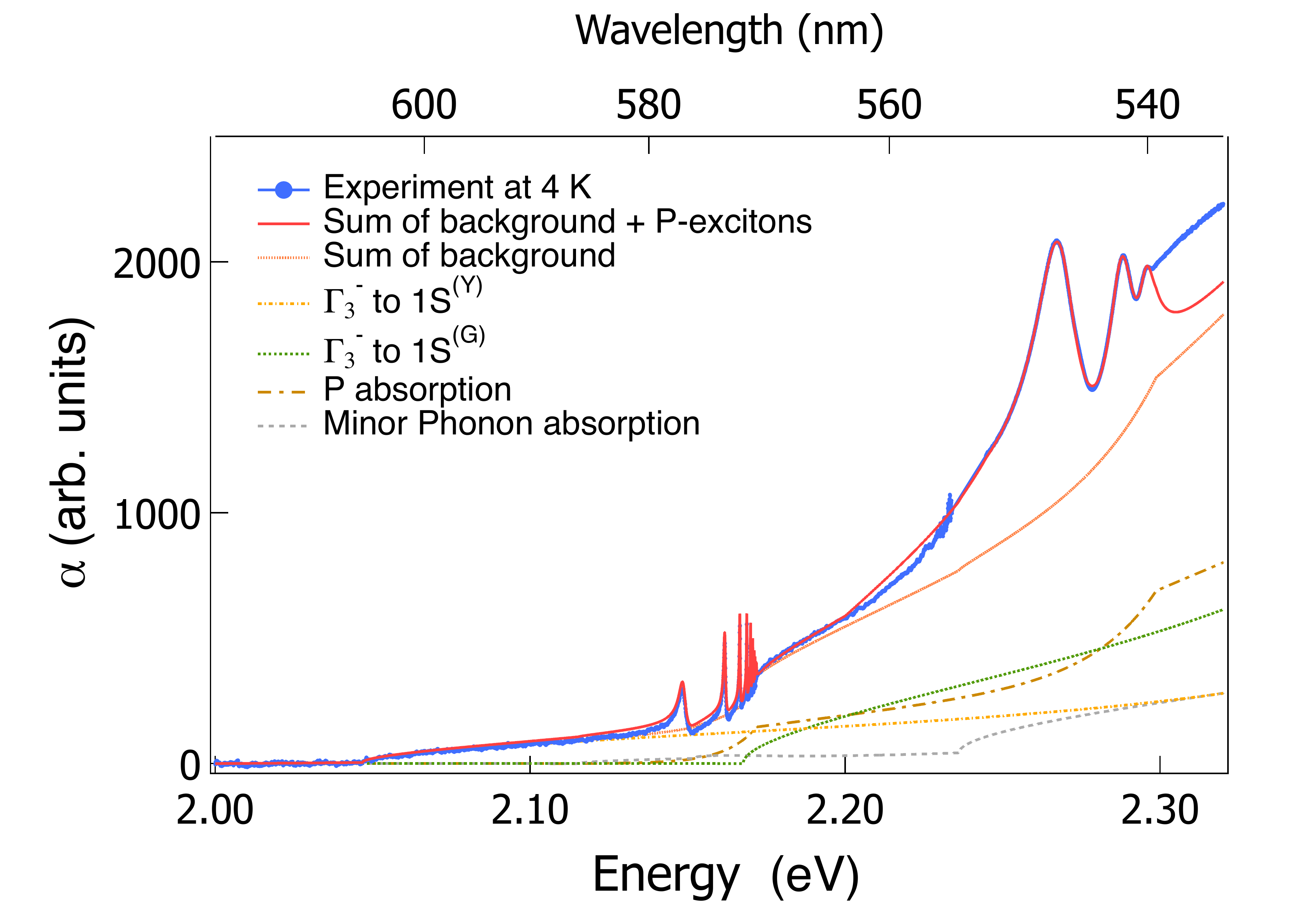}
\caption{Various background absorption spectra and their sum (dotted-orange) with the experimental absorption coefficient spectrum (blue) at 4 K. The red curve shows the theoretically calculated background spectrum summed with the P-exciton resonance, which was reconstructed using Eq. \ref{eq:toyozawa} evaluated with parameter values obtained by fitting to the experimental data.}
\label{Fig7:absco}
\end{figure}

The absorption spectra obtained by the transmission spectroscopy includes the background absorption as well as P-exciton dipole transitions. We use a broadband white light from Ando AQ4303B in the range of 400 – 1800 nm, whose reference spectrum is taken with an 1800 g/mm grating (blaze wavelength 500 nm) in a Princeton Instrument Spectrometer SP-2758. An average power density of the white light at the end of a single-mode fiber is 150 $\mu$W/mm$^2$, which is weak enough not to create the free carriers. According to Ref.~\cite{Heckoetter2018}, the laser induced free carriers can cause the linewidth broadening, the noticeable bandgap shift, relative oscillator strength reduction and the flat continuum background. We have not observed any of these features associated with free carriers in our absorption spectra. 

A recent work \cite{Schone2017} reports that at near-zero temperature, the phonon-assisted absorption spectrum related to the $n$S-exciton is expressed primarily by the ${\Gamma}_3^{-}$, ${\Gamma}_4^{-}$  phonon-assisted transition to the yellow and green 1S excitons. The continuum part of $\alpha_{n\text{S}} (E, T)$ comes mainly from 1S-ortho excitons with emission or absorption of phonons ($E_\text{pn}$). Considering the Elliott's famous formula \cite{Elliott1957, Grun1961}, we can write

\begin{equation}\label{eq:ElittwithT}
\begin{aligned}
&\alpha_{n\text{S}}(E,T) =\\
&A(E,E_{n\text{S}}+\hbar\omega_{3})\sqrt{E-(E_{n\text{S}}+\hbar\omega_{\text{3}})}\left(n_{\omega_{\text{e}}}(T)+1\right)\\ &+A(E,E_{n\text{S}}-\hbar\omega_{3})\sqrt{E-(E_{n\text{S}}-\hbar\omega_{\text{3}})}n_{\omega_{3}}(T), %
\end{aligned}
\end{equation}
for only a dominant phonon $\Gamma_3^-$ with the phonon density in the form of Bose-Einstein distribution
\begin{equation*}
n_{\omega_3}(T)=\left[\exp \left(\hbar \omega_3 / k T\right)-1\right]^{-1},
\end{equation*}
where $A(E, E_{n\text{S}}, \hbar\omega)$ contains material constants, a dipole transition element, and a phonon interaction element with a deformation potential, explicitly written in supplementary materials \cite{Schone2017}.

For example, the expression for the yellow excitons is derived as
\begin{equation}
\begin{split}
\alpha_{n}^{\lambda}(E=\hbar\omega)=&A(E,E_{\text{nS}},\hbar\omega_{\text{pn}})q_{n}^{\lambda},\\
\approxprop&  \sqrt{E-(E_\text{nS}+\hbar\omega_{\text{pn}})},
\end{split}
\end{equation}
with
\begin{equation*}
\begin{aligned}
A(E,E_{\text{ns}},\omega_{\text{pn}} )=&
 \frac{\mathrm{e}^{2}}{\pi^{2} \hbar \rho \varepsilon_{0} n_{R} c a_{\mathrm{b}}^{3}} \frac{M_{\mathrm{y}}}{m_{0}\omega \omega_{\text{pn}}} \frac{\left|p_{78}\right|^{2}}{m_{0}}\left|D_{\lambda ; 68}\left(q_{n}^{\lambda}\right)\right|^{2}\\
 & \times\left|\sum_{n^{\prime}} \frac{\mathcal{S}_{n, n^{\prime}}^{(\mathrm{y}, \mathrm{b})}\left(q_{n}^{\lambda}\right)}{n^{3 / 2}\left(E_{n^{\prime} \mathrm{S}}^{(\mathrm{b})}-\hbar \omega\right)}\right|^{2} \Theta\left(q_{n}^{\lambda}\right), 
\end{aligned}
\end{equation*}
and
\begin{equation*}
\begin{aligned}
q_{n}^{\lambda}(\omega)=\sqrt{\frac{2 M_{\mathrm{y}}}{\hbar^{2}}} \sqrt{\hbar \omega-\hbar \omega_{\text{pn}}-E_{n S}^{(\mathrm{y})}},
\end{aligned}
\end{equation*}
where $\lambda$ is denoted as an available phonon index such as  ${\Gamma}_3^{-}$, ${\Gamma}_4^{-}$ and others; $\hbar\omega_{\text{pn}}$ is phonon energy; $\left|p_{78}\right|$ is the dipole transition element between Bloch states of conduction and valence bands; $\mathcal{S}_{n, n^{\prime}}^{(\mathrm{y}, \mathrm{b})}$ are the overlap factors; $n_{R}$ is the refractive index around the excitation energy; $a_{\mathrm{b}}$ is the Bohr radius of the blue exciton; $M_{\mathrm{y}}$ is the 1s yellow exciton mass.
The values of parameters and material constants in $A$ are chosen from Ref. \cite{Schone2017}.

The bulk phonon-assisted absorption has been typically analyzed by the square root of energy, $\sqrt{E-E_{0}}$, commonly called as the Elliott form \cite{Elliott1957} within the second-order perturbation theory. A recent paper by Sch\"{o}ne $et$  $al.$  \cite{Schone2017} describes an accurate understanding of the dominant ${\Gamma}_3^{-}$ phonon-assisted absorption by 1S ortho-exciton resulting from an arbitrary energy level at a blue exciton level to a yellow exciton level accompanying phonons. Several phonon contributions from theoretical calculations at 4 K are individually plotted in Fig. \ref{Fig7:absco} as a respresentative example. The absorption based on minor phonons such as ${\Gamma}_4^{-}$ ($\hbar\omega_4$ = 82.1 meV) are also taken into account as the Elliott form with the coefficients from the values fitted in Ref. \cite{Schone2017}. The temperature dependence is embedded in phonon occupancy, $n$, given by \cite{Elliott1957}.
Thus $A$ is multiplied by the original phonon-assisted absorption $n+1$
for the phonon creation process and by $n$ for the phonon destruction process.
The red straight line in Fig. \ref{Fig7:absco} displays the overall calculated phonon-assisted absorption spectra, which matches very well with experimental data in most regions.

We also consider the residue of background absorption as P-exciton non-resonant absorption including Urbach tail and continuum absorption.
Specifically, we add the thermal broadening of the Urbach tail in semiconductors, which is given by Ref. \cite{Urbach1953},  
\begin{equation}
\alpha_{\text{Urbach}}(E) = \alpha_{{\text{Urbach}},0} \exp \left[\sigma(T) \frac{\left(E-E_{0}\right)}{k_{B} T}\right],
\end{equation}
and the well-established phonon theory \cite{Dow1972,Rai2013} writes
\begin{equation}
\sigma(T)=\sigma_{0} \frac{2 k_{B} T}{\hbar \omega_{\text{pn}}} \tanh \left(\frac{\hbar \omega_{\text{pn}}}{2 k_{B} T}\right).
\end{equation}

As a result of thorough systematic examinations, the overall background absorption spectra is satisfactorily reproduced for the yellow excitons, from which the yellow $n$P-exciton spectra are confidently isolated in Fig. \ref{Fig2:nPExcitonAbsorption} (a) and their fitted scaling laws are acquired with reliable coincidence to the theoretical expectation in Fig. \ref{Fig3:yellowScaling}. On the other hand, in the case of green excitons, there still exists a non-negligible discrepancy between the experimental absorption coefficient spectrum and the theoretical/fitted background spectrum. Our practical approach to handle the influence of the residual background is to introduce a constant offset term added to Eq. (\ref{eq:toyozawa}) during the fitting process. We found that a constant term was sufficient to accommodate for this residual background and obtain good fitting results for the region of the spectrum shown in Fig. \ref{Fig2:nPExcitonAbsorption} (b), and that additional polynomial terms did not yield any significant improvements to the fit. We think the careful study of absorption process associated with green excitons needs to be done further.

\begin{figure*}[hb]
\includegraphics[width=2\columnwidth]{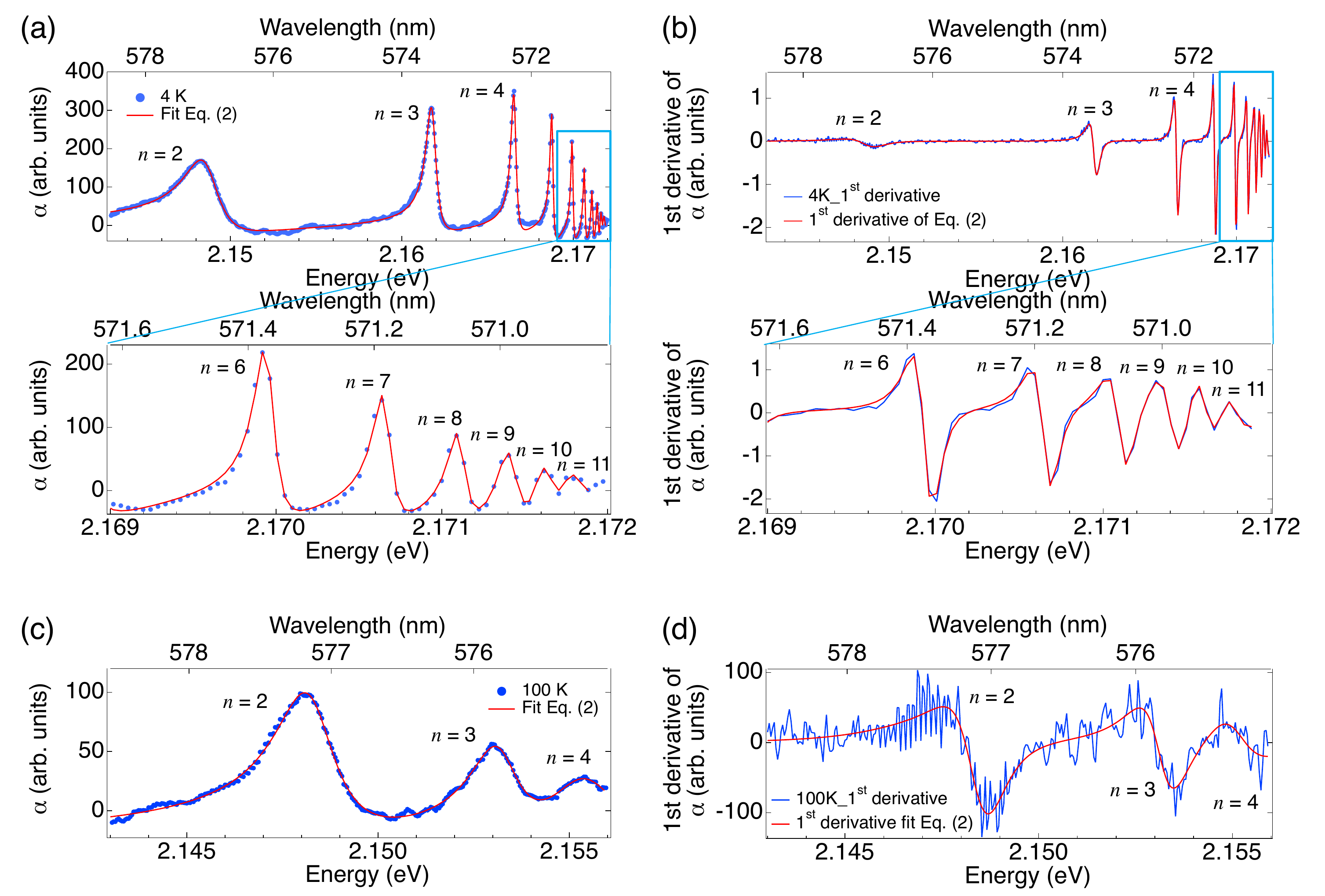}
\caption{Yellow P-exciton resonances. The absorption coefficient spectra $\alpha$ at 4 K ((a)) and 100 K ((c)) and the first derivative of $\alpha$ in energy, $d\alpha/dE$ at 4 K (b) and 100 K (d). Red lines in all figures are the Toyozawa model fits.}
\label{Fig8:PabsZIfit}
\end{figure*}

\subsection*{B. Yellow P-exciton resonances}

In this section, we explicitly explain the fits of Toyozawa's formula to yellow $n$P-exciton resonances in the absorption spectra at two representative temperatures, 4 K (Figs. \ref{Fig8:PabsZIfit}(a) and \ref{Fig8:PabsZIfit}(b)) and 100 K (Figs. \ref{Fig8:PabsZIfit}(c) and \ref{Fig8:PabsZIfit}(d)). This work is the basis for the extraction of optical properties presented in Figs. \ref{Fig3:yellowScaling} and \ref{Fig5:greenScaling} from Toyozawa's exciton model.  In order to access higher $n$-exciton states, we inspect the absorption coefficient spectra and their energy derivatives. Experimental data are presented in blue symbols and the model fits are shown in red straight lines in Fig. \ref{Fig8:PabsZIfit}. 
By and large, the Toyozawa's asymmetric Lorentzian fits work extremely well both in the original and the first energy derivative of the absorption spectra. At 4 K (Fig. \ref{Fig8:PabsZIfit}(a)), it is clear that the peaks of $n <$ 8-9 are asymmetric Lorentzian shaped, whereas the higher $n$ peaks are more ambiguously shaped, partly because they are convoluted with the spectral resolution of our monochromator. As temperature increases, despite the spectral linewidth broadening, the peaks look still closer to asymmetric Lorentzian as can be seen in the spectra at 100 K (Figs. \ref{Fig8:PabsZIfit}(c) and \ref{Fig8:PabsZIfit}(d)). The analysis of the first derivative confirms the values of the parameters for exciton binding energies, FWHM, peak areas and the asymmetry of the peaks.\\

\subsection*{C. Photoluminescence of the yellow exciton series in \ce{Cu2O}}

\begin{figure}[htbp]
\includegraphics[width=0.95\columnwidth]{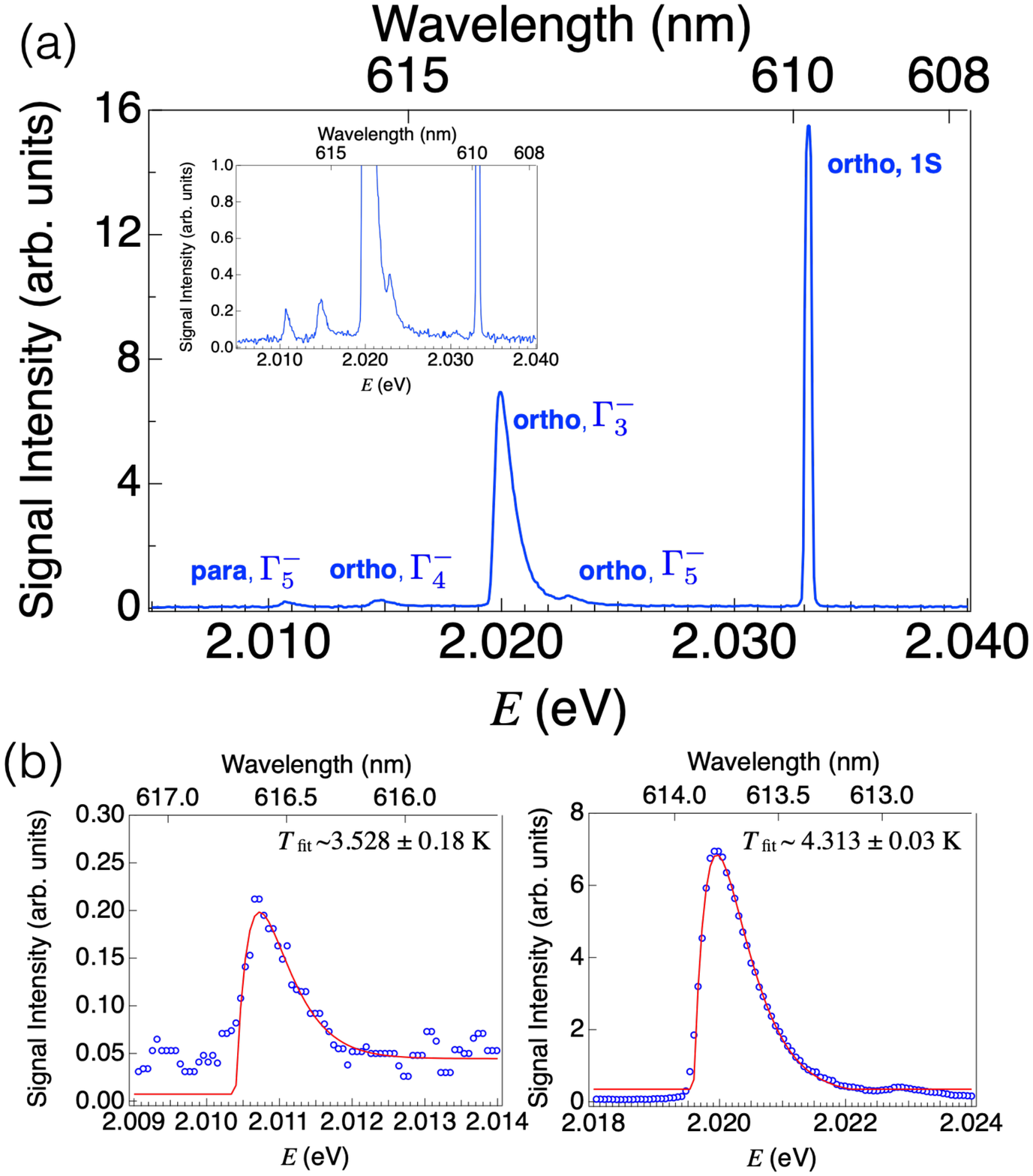}
\caption{(a) Photoluminescence spectrum at 4 K obtained using a 520 nm laser. The inset shows the enlarged PL spectra for the clear phonon side bands. (b) Maxwell-Boltzmann distribution fits to the $\Gamma_{5}^{-}$ para-exciton near 2.011 eV (left) and the $\Gamma_{3}^{-}$ ortho-exciton near 2.020 eV. 
}
\label{Fig9:PL}
\end{figure}

The yellow 1S exciton is dipole-forbidden; however, the direct luminescence of the 1S ortho-exciton (triplet) is still the most dominant radiative recombination process via electric quadrupole transition \cite{Takahata2018}. Since PL spectroscopy enables us to assess the 1S ortho-exciton related peaks, both direct and phonon side bands, we perform PL measurements in a transmission configuration with the excitation of a 520 nm semiconductor laser diode, and collect the luminescence signal in forward direction.  
The representative spectrum at $\sim$4 K is obtained around the 1S yellow-exciton energy level (Fig. \ref{Fig9:PL}(a)). Similarly to previous papers \cite{Takahata2018}, the 1S ortho-exciton peak appears at 2.033 eV ($\sim$609.8 nm), with several phonon-assisted signal peaks. As temperature goes up,  the 1S ortho-exciton energy exhibits a red-shift (Fig. \ref{Fig1:PhononAssistedAbsorption}(c)). 
The binding energy of 1S ortho-exciton from $E_{\text{g}}$ in Fig. \ref{Fig2:nPExcitonAbsorption} is about 139.4 meV at 4 K, with a slightly larger value of 141.3 meV at 100 K. 

We draw the attention to the fact that the ortho-exciton peak ($\Gamma_{3}^{-}$-phonon replica) is the largest peak, being the dominant phonon in \ce{Cu2O} and providing evidence of the interaction between the $\Gamma_{3}^{-}$ phonon and 1S exciton. Furthermore, this peak exhibits an energy shift identical to the 1S ortho-exciton across the observed temperature ranges, keeping the phonon energy at a constant value of 13.6 meV. We apply the Maxwell-Boltzmann distribution to the phonon-assisted peak for extracting the peak energy at effective temperature $T_{\mathrm{eff}}$ in the intensity $I(E)$ expression given by \cite{Takahata2018},
\begin{equation}I(E) \propto \sqrt{E-E_{\mathrm{i}}} \exp \left[-\left(E-E_{\mathrm{i}}\right) /\left(k_{\mathrm{B}} T_{\mathrm{eff}}\right)\right].
\end{equation}

Our PL spectra also supports our crystal being of high quality, with the clear observation of the para-exciton signal ($\Gamma_{5}^{-}$-phonon replica) at 2.01066 eV ($\sim$616.635 nm) (Fig.~\ref{Fig9:PL}(a),inset). Furthermore, the relative intensity strengths of the quadrupole resonance and the $\Gamma_{5}^{-}$ para-exciton peak with respect to the $\Gamma_{3}^{-}$ ortho-exciton peak are 2.2 and 0.08, which is comparable to the values of 3.9 and 0.06 from the PL spectra of the high-purity \ce{Cu2O} crystal taken at 2 K \cite{YoshiokaPhD}.

We further confirm the very high-purity of our sample from the Maxwell-Boltzmann (MB) distribution fits to the 1s, ortho-exiton ($\Gamma_{3}^{-}$-phonon replica) and the 1s, para-exciton ($\Gamma_{3}^{-}$-phonon replica) luminescence signals shown in Fig.~\ref{Fig9:PL}(b). Both peaks match very well with the MB fitting from which we can find the effective temperatures. While the ortho-excitons reside at around 4.31 $\pm$ 0.03 K, the very long-lived para-excitons are indeed lower at 3.53 $\pm$ 0.18 K, which would be the lattice temperature.

\clearpage
\end{document}